\documentclass[journal]{IEEEtran}

\usepackage{pgfplots}
\usepackage[bookmarks,colorlinks]{hyperref}
\usepackage[linesnumbered,ruled,lined]{algorithm2e}
\usepackage{enumitem}
\usepackage{algpseudocode}
\usetikzlibrary{shapes.multipart,intersections}
\usepackage{cite}
\usepackage{amsmath,amssymb,amsfonts,amsthm,steinmetz}
\usepackage{graphicx}
\usepackage{mathrsfs}  
\usepackage{textcomp}
\usepackage{acronym}
\usepackage{xcolor}
\usepackage{upgreek,xspace}

\usepackage{comment}

\usepackage{tikz}
\usetikzlibrary{calc}
\makeatletter
\newcommand{\gettikzxy}[3]{%
  \tikz@scan@one@point\pgfutil@firstofone#1\relax
  \edef#2{\the\pgf@x}%
  \edef#3{\the\pgf@y}%
}
\makeatother

\usepackage[draft]{todonotes}

\usepackage{esvect}

\usepackage{booktabs}

\usepackage{times}
\usepackage{bm}
\usepackage{amsmath}
\usepackage{amssymb}
\usepackage{stmaryrd}
\usepackage{babel}
\usepackage{graphics, graphicx}
\usepackage{xcolor}
\usepackage{gensymb}
\usepackage{cite}
\usepackage{enumitem}
\usepackage{url}

\begin{document}

\title{Experimental Multiport-Network Parameter Estimation for a Dynamic Metasurface Antenna}

\author{Jean~Tapie~and~Philipp~del~Hougne,~\IEEEmembership{Member,~IEEE}
\thanks{
J.~Tapie and P.~del~Hougne are with Univ Rennes, CNRS, IETR - UMR 6164, F-35000, Rennes, France (e-mail: \{jean.tapie; philipp.del-hougne\}@univ-rennes.fr). P.~del~Hougne is also with Aalto University, Department of Electronics and Nanoengineering, 02150 Espoo, Finland.
}
\thanks{\textit{(Corresponding Author: Philipp del Hougne.)}}
\thanks{This work was supported in part by the Nokia Foundation (project 20260028), the ANR France 2030 program (project ANR-22-PEFT-0005), the ANR PRCI program (project ANR-22-CE93-0010), the French Defense Innovation Agency (project 2024600), the European Union's European Regional Development Fund, and the French region of Brittany and Rennes Métropole through the contrats de plan État-Région program (projects ``SOPHIE/STIC \& Ondes'' and ``CyMoCoD'').}
}

\maketitle

\begin{abstract}
Most use cases of reconfigurable antennas require an accurate forward model mapping configuration to radiated field (and reflections at feeds). Emerging dynamic metasurface antennas (DMAs) confront the conventional approach of extracting such a model from a numerical simulation with multiple challenges. 
\textit{First}, the cost of accurately simulating an intricate and electrically large DMA architecture might be prohibitive.
\textit{Second}, the model-reality mismatch due to fabrication inaccuracies might be substantial, especially at higher frequencies and for DMA architectures leveraging strong inter-element mutual coupling (MC) to maximize their tunability. 
These considerations motivate an \textit{experimental} parameter estimation for DMA forward models. 
The main challenge lies in the forward model's non-linearity due to inter-element MC.
Multiport network theory (MNT) can accurately capture MC but the MC parameters cannot be measured directly.
In this article, we demonstrate the experimental estimation of a high-accuracy proxy MNT model for a 19-GHz DMA with 7 feeds and 96 elements, where all feeds and elements are strongly coupled via a chaotic cavity. 
For a given DMA configuration and excitation, our proxy MNT model predicts the reflected field at the feeds and the radiated field with accuracies of 40.3~dB and 37.7~dB, respectively. A simpler, MC-unaware benchmark model only achieves 2.6~dB and 3.3~dB, respectively.
We systematically examine the influence of the number of feeds and measured DMA configurations on the model accuracy, motivating the inclusion of ``auxiliary calibration feeds'' to facilitate the parameter estimation when the intended DMA operation is limited to a single feed.
Finally, we measure DMA configurations optimized based on our proxy MNT model.
\end{abstract}


\section{Introduction}\label{sec_Introduction}

Large-scale multi-element antenna arrays enable digital beamforming but are confronted with cost and power consumption challenges due to the large number of required radiofrequency (RF) chains. This concern motivates the exploration of hybrid analog/digital beamforming based on antenna architectures with wave-domain flexibility. Originally envisioned hybrid beamforming architectures consist in connecting a reduced number of RF chains via a reconfigurable analog combining circuit to a large-scale multi-element antenna array~\cite{zhang2005variable,venkateswaran2010analog,roh2014millimeter,han2015large,gong2020rf}. Recent advances in metasurface technology enable a direct integration of reconfigurable analog combining into the antenna array. A prominent emerging example is the dynamic metasurface antenna (DMA) which is an ultrathin device coupling a few feeds to hundreds of reconfigurable radiating elements via planar microstrips or cavities~\cite{sleasman2015dynamic,yurduseven2018dynamically,yoo2018enhancing,sleasman2020implementation,del2020learned,boyarsky2021electronically,shlezinger2021dynamic,jabbar202460,yven2025end}. The concept of wave-domain reconfigurability is also found in related antenna technologies like electronically steerable passive array radiator (ESPAR) antennas~\cite{harrington1978reactively,schlub2003seven,sun2004fast,kawakami2005electrically,lu2005dielectric,luther2012microstrip,movahedinia2018} and reconfigurable pixel-based antennas~\cite{Flaviis_PixelAntenna,rodrigo2012frequency,MURCH_TAP_PixelAntenna}.  
Moreover, transceiver architectures based on transmit-arrays, reflect-arrays or reconfigurable intelligent surfaces (RISs) leverage wave-domain programmability~\cite{sievenpiper2002tunable,sievenpiper2003two,kamoda201160,clemente20121,CQW14}, although typically in a less compact form-factor than DMAs, ESPARs or reconfigurable pixel-based antennas. 
For concreteness, we focus on DMAs in this article, but our discussion and findings extend to any reconfigurable antenna architecture involving tunable lumped elements.

A prerequisite for the system-level deployment of a DMA is the availability of an accurate forward model. A forward model predicts the DMA's radiated field, as well as the reflected field at the DMA's feeds\footnote{Knowing the reflected field at the feeds is important to assess (i) the reflected power impinging on the RF chains, (ii) the isolation between feeds, and (iii) the correlations between the thermal noise at different feeds~\cite{wedge2002noise}.}, as a function of the DMA's configuration and excitation. Based on the forward model, the DMA configuration can be optimized in line with system requirements. Theoretical studies typically assume that the forward model is known perfectly and oftentimes neglect important electromagnetic effects like inter-element mutual coupling (MC). Antenna-level works typically assume that a suitable forward model can be extracted from an accurate numerical full-wave simulation of the known DMA architecture. However, this conventional approach is confronted with challenges:
\begin{enumerate}[label=(\roman*)]
    \item The DMA is an electrically large structure with many intricate, electrically small features. Even if all details (geometry and material composition) are known perfectly, the computational cost of a full-wave simulation might be (prohibitively) high.
    \item The risk of model-reality mismatch is significant due to fabrication inaccuracies, especially at higher frequencies and for DMAs with strong inter-element MC~\cite{xu2022extreme,xu2022wide}. (Recently, strong inter-element MC is emerging as a means of maximizing a DMA's wave-domain flexibility with a fixed number of tunable elements~\cite{prod2024mutual,MCbenefitsDMA}.)
\end{enumerate}

Existing experimental demonstrations of model-based control of DMAs are typically based on reduced-order MC-unaware models and limited to DMA prototypes where inter-element MC is deliberately mitigated~\cite{boyarsky2021electronically,jabbar202460}. However, as mentioned, strong inter-element MC is desirable to maximize the DMA's wave-domain flexibility~\cite{prod2024mutual,MCbenefitsDMA}.
MC-aware reduced-order models based on a discrete-dipole approximation (DDA)~\cite{yoo2019analytic,yoo2020analytic} or multiport-network theory (MNT)~\cite{williams2022electromagnetic,grbic2025,salmi2025optimization,ramirez2025metasurface,MultiPortDMA_EuCAP2026} accurately describe the electromagnetic interactions between the DMA's elements. However, the typical reliance on analytically or numerically estimated parameters as in~\cite{yoo2019analytic,yoo2020analytic,williams2022electromagnetic,grbic2025,salmi2025optimization,ramirez2025metasurface} risks once again significant model-reality mismatch due to fabrication inaccuracies.
Indeed, the only work on model-based control of a DMA with significant inter-element MC that we are aware of uses numerically extracted MC-aware model parameters and observes notable discrepancies between the model prediction and the experimental reality~\cite{xu2022wide}; we emphasize that this mismatch occurs despite the fact that the inter-element MC strength was deliberately limited in~\cite{xu2022wide} to reduce the vulnerability to fabrication inaccuracies.

To prevent any model-reality mismatch of MC-aware forward models for DMAs with significant inter-element MC, it is desirable to experimentally estimate the model parameters. However, inter-element MC parameters cannot be measured directly: The DMA's elements are not connectorized and, in any case, their number exceeds the number of ports of a typical vector network analyzer by at least one order of magnitude. Encouragingly, recent work in the realm of RIS-parametrized wireless channels demonstrated the experimental estimation of accurate proxy MNT models~\cite{sol2023experimentally,del2025experimental,del2025ambiguity,del2025rank}. The attribute ``proxy'' highlights that the parameters cannot be identified unambiguously (because the conditions for lifting all ambiguities outlined in~\cite{del2025virtual_2p0} are not met), but this is operationally irrelevant because the end-to-end channels are accurately predicted. Importantly, the number of model parameters is independent of the complexity of the mechanisms that mediate the coupling between the entities described as ports within MNT. As we elaborate in Sec.~\ref{sec_theory}, the MNT system models for RIS-parametrized channels and DMAs are closely related. The pivotal requirement is that the tunable elements are lumped, i.e., very small compared to the wavelength. Preliminary work for the present article described in~\cite{MultiPortDMA_EuCAP2026} confirmed the validity of this requirement for our K-band DMA prototype (described in Sec.~\ref{subsec_prototype}).

Our contributions are summarized as follows:
\begin{enumerate}
    \item We propose a procedure for experimentally estimating the parameters of a proxy MNT model for DMAs with inter-element MC.
    \item We experimentally validate our MNT parameter estimation based on the first multi-port chaotic-cavity-based DMA prototype. (The chaotic cavity induces strong MC between all feeds and tunable elements.)
    \item We systematically examine the influence of the number of feeds and the number of measurements (with distinct DMA configurations) on the proxy MNT model's accuracy. We introduce the idea of ``auxiliary calibration feeds'' (feeds that are only used for parameter estimation and then open-circuited during operation).
    \item We benchmark the MC-aware MNT model's accuracy against that of an MC-unaware  model.
    \item We optimize DMA configurations based on our proxy MNT model (as well as the MC-unaware benchmark model) for various beamforming tasks, and we measure the corresponding radiated fields, as well as the corresponding reflected fields at the feeds.
\end{enumerate}

\textit{Organization:}
In Sec.~\ref{sec_theory}, we define our MNT system model, as well as an MC-unaware benchmark model.
In Sec.~\ref{sec_algorithm}, we describe our algorithm to estimate proxy MNT model parameters, as well as an algorithm to estimate proxy parameters for the MC-unaware benchmark model.
In Sec.~\ref{sec_ExpValidation}, we describe our DMA prototype, our measurement setup and procedure, a few system model sanity checks, and a systematic analysis of the accuracy achieved with the algorithms from Sec.~\ref{sec_algorithm}.
In Sec.~\ref{sec_optimization}, we report on measurements of DMA configurations optimized based on our calibrated model from Sec.~\ref{sec_ExpValidation} for single-user and dual-user focusing.
In Sec.~\ref{sec_Conclusion}, we close with a brief summary and outlook.

\textit{Notation:}
$\mathbf{A}_\mathcal{BC}$ denotes the block of the matrix $\mathbf{A}$ whose row [column] indices are in the set $\mathcal{B}$ [$\mathcal{C}$].
$\mathbb{B}$ denotes the binary set $\{0,1\}$.
$\mathcal{B}_i$ is the singleton containing the $i$th entry of $\mathcal{B}$.
$\mathbf{I}_a$ is the $a\times a$ identity matrix.
$\mathbf{A}^\top$ denotes the  transpose of $\mathbf{A}$.
$\mathbf{A}^\dagger$ denotes the conjugate transpose of $\mathbf{A}$.
$\mathbf{A}^+$ denotes the Moore–Penrose pseudoinverse of $\mathbf{A}$.
$\mathrm{vec}(\mathbf{A})$ denotes the column-stacking vectorization of $\mathbf{A}$.
$\mathbf{0}$ denotes a zero vector or zero matrix whose size is implied by the context.
$\mathbf{1}_a$ denotes an $a$-element vector whose entries are unity.
$\odot$ denotes the Khatri--Rao (column-wise Kronecker) product.
$\jmath=\sqrt{-1}$ denotes the imaginary unit.

Throughout this paper, we work with scattering parameters; our results can be also expressed in equivalent representations like impedance and admittance parameters.

\section{System Model}
\label{sec_theory}

A DMA can be defined as an ultrathin structure patterned with $N_\mathrm{M}$ programmable meta-elements on the front and $N_\mathrm{F}$ feeds on the back. During transmission, the RF chains connected to the feeds generate an input signal $\mathbf{x}\in\mathbb{C}^{N_\mathrm{F}}$ that excites the DMA via its feeds. The signal couples from the feeds to the meta-elements in a manner that depends on the specific DMA architecture; for example, the coupling mechanism can rely on microstrips~\cite{sleasman2015dynamic} or cavities~\cite{sleasman2020implementation}. The superposition of the signals leaking out of the DMA via the meta-elements forms the radiated field. The meta-elements are equipped with independently tunable components that control the signal leakage; the configuration of the tunable components also generally affects the coupling between feeds and meta-elements. So-called ``beyond-diagonal'' DMAs (BD-DMAs) are additionally equipped with tunable components in the coupling structure between feeds and meta-elements to further increase the wave-domain flexibility~\cite{prod2025beyond}. We limit our discussion for simplicity to a DMA operated in transmission, but the reversed version of the described transmission scenario directly yields the reception scenario (with signal-generating RF chains replaced by signal-detecting RF chains).

We now further abstract the DMA as a radiating structure excited by a wavefront $\mathbf{x}$ via $N_\mathrm{F}$ monomodal ports and parametrized by $N_\mathrm{M}$ tunable lumped elements. Each tunable lumped element can be interpreted as an additional ``virtual'' lumped port to the structure that is terminated by a tunable load. Overall, the DMA thus consists of a radiating static structure with $N_\mathrm{F}+N_\mathrm{M}$ monomodal ports, of which $N_\mathrm{M}$ ports are terminated by tunable loads. We emphasize the general nature of this description that makes no assumption about the details of the mechanisms via which the static structure couples the $N_\mathrm{F}+N_\mathrm{M}$ ports (e.g., waveguides, cavities) nor about the meta-element design. The description equally applies to conventional DMAs and BD-DMAs, and also to any other reconfigurable antenna architecture whose reconfigurability originates from tunable lumped elements (such as ESPARs or reconfigurable pixel-based antennas).

To complete the description of the DMA as a radiating structure, a formalism to describe its radiated field is required. Recent theoretical works on reconfigurable antennas used embedded-element patterns (EEPs)~\cite{grbic2025,salmi2025optimization}; alternatively, the antenna literature offers the concept of ``radiation ports'' (defined in bases such as spherical harmonics~\cite{hansen1988spherical}); radiation ports are generally assumed to not reflect any waves and to not feature any mutual coupling between them. In measurements, the radiated field is sampled at a discrete set of grid points, typically with a ``virtual antenna array'' (VAA) that probes the radiated field by mechanically displacing a probe or an array of probes (and/or, equivalently, rotating the antenna under test). EEPs or scattering into radiation ports can then be deduced from the VAA measurements. In this paper, we work directly in the VAA basis, deferring transformations to other bases to future work. With $N_\mathrm{G}$ grid points, the VAA has $N_\mathrm{G}$ lumped monomodal ports.

\begin{figure}
    \centering
    \includegraphics[width=\columnwidth]{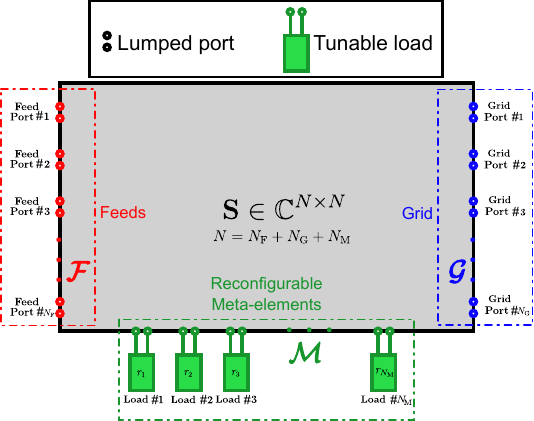}
    \caption{MNT system model for a DMA with $N_\mathrm{F}$ feeds and $N_\mathrm{M}$ reconfigurable meta-elements; the radiated field is probed by a VAA with $N_\mathrm{G}$ grid points.}
    \label{Fig1}
\end{figure}

Now, we can formulate the MNT system model displayed in Fig.~\ref{Fig1}. Our system consists of a static $N$-port structure, where $N=N_\mathrm{F}+N_\mathrm{M}+N_\mathrm{G}$, and a set of tunable individual loads terminating $N_\mathrm{M}$ of the static structure's ports. The static structure is linear, passive, time-invariant, and reciprocal, and characterized by a scattering matrix $\mathbf{S}\in\mathbb{C}^{N \times N}$ that naturally admits the following partition:
\begin{equation}
\mathbf{S} =\begin{bmatrix}
\mathbf{S}_\mathcal{FF} & \mathbf{S}_\mathcal{FM} & \mathbf{S}_\mathcal{FG}\\
\mathbf{S}_\mathcal{MF} & \mathbf{S}_\mathcal{MM}  & \mathbf{S}_\mathcal{MG} \\
\mathbf{S}_\mathcal{GF} & \mathbf{S}_\mathcal{GM}  & \mathbf{S}_\mathcal{GG} 
\end{bmatrix}
\in \mathbb{C}^{N \times N},
\label{eq1}
\end{equation}
where $\mathcal{F}$ denotes the set of port indices associated with the DMA's feeds, $\mathcal{G}$ denotes the set of port indices associated with the VAA grid, and $\mathcal{M}$ denotes the set of port indices associated with the reconfigurable meta-elements.
Meanwhile, the $i$th individual load is characterized by its reflection coefficient $r_i\in\mathbb{C}$; the ensemble of the $N_\mathrm{M}$ individual loads is characterized by a diagonal scattering matrix: 
\begin{equation}
    \mathbf{\Phi}= \mathrm{diag} (\mathbf{r})\in\mathbb{C}^{N_{\mathrm{M}}\times N_{\mathrm{M}}},
\end{equation}
where $\mathbf{r} = [r_{1},...,r_{N_{\mathrm{M}}}]$.

Our two quantities of interest, namely the radiated signal $\mathbf{y}_\mathrm{G}\in\mathbb{C}^{N_\mathrm{G}}$ at the VAA grid points and the reflected signal $\mathbf{y}_\mathrm{F}\in\mathbb{C}^{N_\mathrm{F}}$ at the feeds, depend on both $\mathbf{x}$ (the signal exciting the DMA via its feeds) and $\mathbf{r}$:
\begin{subequations}
\label{eq_LinSysResp}
\begin{align}
\mathbf{y}_\mathrm{G} &= \mathbf{T}(\mathbf{r})\,\mathbf{x}, \\
\mathbf{y}_\mathrm{F} &= \mathbf{R}(\mathbf{r})\,\mathbf{x},
\end{align}
\end{subequations}
where $\mathbf{T}\in\mathbb{C}^{N_\mathrm{G} \times N_\mathrm{F}}$ and $\mathbf{R}\in\mathbb{C}^{N_\mathrm{F} \times N_\mathrm{F}}$ denote, respectively the transmission matrix for signals transmitted from feeds to VAA grid points and the reflection matrix for signals reflected at the feeds, respectively. Standard MNT~\cite{anderson_cascade_1966,ha1981solid,prod2024efficient} relates $\mathbf{T}$ and  $\mathbf{R}$ to $\mathbf{S}$ and $\mathbf{\Phi}$:
\begin{subequations}
\label{eq_MNT}
\begin{equation}
    \begin{aligned}
\mathbf{T} &= {\mathbf{S}}_\mathcal{GF} +{\mathbf{S}}_\mathcal{GM} \left( \mathbf{I}_{N_\mathrm{M}} - \mathbf{\Phi}\,{\mathbf{S}}_\mathcal{MM}  \right)^{-1} \mathbf{\Phi} \, {\mathbf{S}}_\mathcal{MF}.
    \end{aligned}
    \label{eq_MNT_T}
\end{equation}
\begin{equation}
\begin{aligned}
    \mathbf{R} &= {\mathbf{S}}_\mathcal{FF} +{\mathbf{S}}_\mathcal{FM} \left( \mathbf{I}_{N_\mathrm{M}} - \mathbf{\Phi}\,{\mathbf{S}}_\mathcal{MM}  \right)^{-1} \mathbf{\Phi} \, {\mathbf{S}}_\mathcal{MF},
    \end{aligned}
    \label{eq_MNT_R}
\end{equation}
\end{subequations}
We emphasize that we made no assumption about the separation between DMA and VAA. We further note that $\mathbf{S}_\mathcal{GG}$ plays no operational role in (\ref{eq_MNT}). 
Our definitions of the scattering matrices $\mathbf{S}$ and $\mathbf{\Phi}$ use a reference impedance of 50~$\Omega$ at all ports, and we assume that the feeds and VAA probe are connected to measurement equipment whose ports are matched to 50~$\Omega$, which is the case for typical vector network analyzers (VNAs) as well as signal generators and detectors.

So far, we have established a wave-domain system model. We still need to account for the mapping from the control vector for the DMA configuration to $\mathbf{r}$, which is trivially known in theoretical work but an additional challenge in experiments.
For concreteness and to avoid overly complicated generic notation, we specialize this discussion to our DMA prototype with identical and independently controllable elements with 1-bit programmability. Let $\alpha\in\mathbb{C}$ and $\beta\in\mathbb{C}$ denote the two possible values of the reflection coefficient that the loads can have. Then, the control vector $\mathbf{b}\in\mathbb{B}^{N_\mathrm{M}}$ maps to $\mathbf{r}$ as follows:
\begin{equation}
    \mathbf{r} = \alpha\mathbf{1}_{N_\mathrm{M}} + (\beta-\alpha)\mathbf{b}.
    \label{eq_b2r}
\end{equation}
We emphasize that the mapping from $\mathbf{b}$ to $\mathbf{r}$ can in general be much more complicated than the simple affine relation in (\ref{eq_b2r}). With more than two possible states, the mapping is generally not affine. Moreover, the control circuit can create correlations between the entries of $\mathbf{r}$, for instance, due to a limited power supply. Furthermore, it is in principle conceivable that the tunable elements differ regarding the characteristics of their states, and even the number of accessible states.

\begin{figure}
    \centering
    \includegraphics[width=\columnwidth]{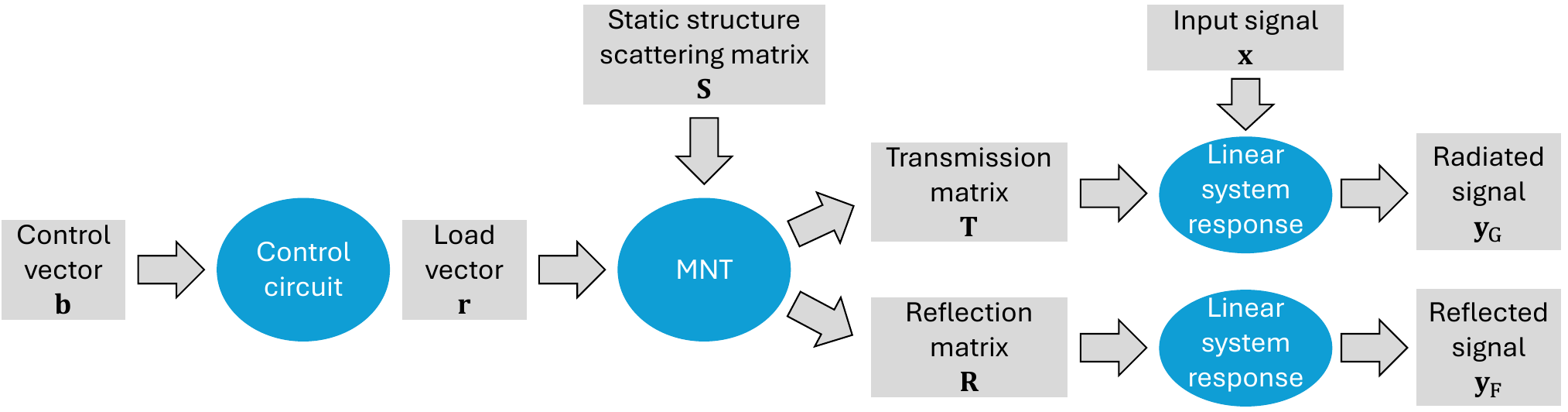}
    \caption{Overview of mapping from control vector $\mathbf{b}$ and input signal $\mathbf{x}$ to radiated signal $\mathbf{y}_\mathrm{G}$ and reflected signal $\mathbf{y}_\mathrm{F}$.  ``Control circuit'' refers to (\ref{eq_b2r}), ``MNT'' refers to (\ref{eq_MNT}), ``Linear system response'' refers to (\ref{eq_LinSysResp}).}
    \label{Fig2}
\end{figure}

Altogether, as summarized in Fig.~\ref{Fig2}, in this section, we have established a system model based on MNT for mapping the control vector $\mathbf{b}$ and the excitation signal $\mathbf{x}$ to the radiated field $\mathbf{y}_\mathrm{G}$ and the field reflected at the feeds $\mathbf{y}_\mathrm{F}$. The required parameters are $\alpha$ and $\beta$, as well as the entries of $\mathbf{S}_\mathcal{FF}$, $\mathbf{S}_\mathcal{FM}$, $\mathbf{S}_\mathcal{MM}$, $\mathbf{S}_\mathcal{GF}$, and $\mathbf{S}_\mathcal{GM}$. While these parameters are perfectly known in theoretical work, our goal in this paper is to estimate them experimentally to avoid any model-reality mismatch due to fabrication inaccuracies, as well as the computational cost of numerical full-wave simulations of the DMA.

It is instructive to compare (\ref{eq_MNT}) with the MNT model for RIS-parametrized end-to-end wireless MIMO channels in~\cite{sol2023experimentally,del2025experimental,del2025ambiguity,del2025rank}. A direct analogy with (\ref{eq_MNT_T}) is apparent, where the DMA's feeds correspond to the transmitting antennas' ports, the meta-elements' tunable lumped elements correspond to RIS elements' tunable lumped elements, and the VAA ports correspond to the receiving antennas' ports. This analogy reiterates the interpretation of a DMA as a device that compactly integrates both an RIS and the transmitting antennas; for a base station use case, this compactness of the DMA is very attractive.
At the same time, we note that the mathematical structure of (\ref{eq_MNT_T}) and (\ref{eq_MNT_R}) is very similar, and that both share the term $\left( \mathbf{I}_{N_\mathrm{M}} - \mathbf{\Phi}\,{\mathbf{S}}_\mathcal{MM}  \right)^{-1} \mathbf{\Phi} \, {\mathbf{S}}_\mathcal{MF}$.

Before closing this section, we define an MC-unaware benchmark model. This benchmark model simply consists in assuming $\mathbf{S}_\mathcal{MM}=\mathbf{0}$, which yields

\begin{subequations}
\label{eq_noMC}
\begin{equation}
    \begin{aligned}
\mathbf{T}^\mathrm{noMC} &= {\mathbf{S}}_\mathcal{GF} +{\mathbf{S}}_\mathcal{GM}  \mathbf{\Phi} \, {\mathbf{S}}_\mathcal{MF}.
    \end{aligned}
    \label{eq_noMC_T}
\end{equation}
\begin{equation}
\begin{aligned}
    \mathbf{R}^\mathrm{noMC} &= {\mathbf{S}}_\mathcal{FF} +{\mathbf{S}}_\mathcal{FM}  \mathbf{\Phi} \, {\mathbf{S}}_\mathcal{MF},
    \end{aligned}
    \label{eq_noMC_R}
\end{equation}
\end{subequations}

\section{MNT Parameter Estimation Algorithm}
\label{sec_algorithm}

In this section, we describe our algorithm to estimate a set of parameters such that we can accurately map $\mathbf{b}$ and $\mathbf{x}$ to $\mathbf{y}_\mathrm{G}$ and $\mathbf{y}_\mathrm{F}$ based on our MNT model (see Fig.~\ref{Fig2}). 

\subsection{Background}
\label{subsec_backgroundAlgorithm}

As mentioned, experimental measurements cannot unambiguously retrieve the MNT model parameters\footnote{The method for unambiguously estimating $\mathbf{S}$ in~\cite{del2025virtual_2p0} assumes that the virtual ports associated with tunable lumped elements can be terminated by three distinct known individual loads as well as coupled loads. The results in~\cite{Montpellier1} suggest that two distinct individual loads can be enough because the coupled loads provide a distinct third termination of the virtual ports. In any case, our DMA has no coupled loads and we assume its two distinct loads to be unknown, precluding an unambiguous estimation of $\mathbf{S}$.   } but they can identify a set of proxy parameters (denoted with a tilde: $\tilde{\alpha}$,\,$\tilde{\beta}$,\,$\tilde{\mathbf{S}}_\mathcal{FF},\,\tilde{\mathbf{S}}_\mathcal{FM},\,\tilde{\mathbf{S}}_\mathcal{MM},\,\tilde{\mathbf{S}}_\mathcal{GF},\,\tilde{\mathbf{S}}_\mathcal{GM}$) that accurately map $\mathbf{b}$ and $\mathbf{x}$ to $\mathbf{y}_\mathrm{G}$ and $\mathbf{y}_\mathrm{F}$ based on our MNT model. To avoid any model-reality mismatch and any need for numerical full-wave simulations, we do not assume to know any of the proxy parameters a priori. We emphasize that we thus do \textit{not} assume to know the two possible reflection coefficients $\alpha$ and $\beta$ of the tunable lumped elements; thereby, we avoid reliance on being able to obtain frequency-resolved, complex-valued reflection coefficients for the tunable lumped elements from their data sheet.

Accounting for reciprocity ($\tilde{\mathbf{S}}_\mathcal{FF}=\tilde{\mathbf{S}}_\mathcal{FF}^\top$ and $\tilde{\mathbf{S}}_\mathcal{MM}=\tilde{\mathbf{S}}_\mathcal{MM}^\top$, and $\tilde{\mathbf{S}}_\mathcal{FM}=\tilde{\mathbf{S}}_\mathcal{MF}^\top$), the total number of unknown complex-valued parameters is thus
\begin{equation}
\begin{aligned}
n_\mathrm{u}
= 2
&+ \frac{N_\mathrm{F}(N_\mathrm{F}+1)}{2}
+ N_\mathrm{F}N_\mathrm{M}
+ \frac{N_\mathrm{M}(N_\mathrm{M}+1)}{2}\\
&+ N_\mathrm{G}N_\mathrm{F}
+ N_\mathrm{G}N_\mathrm{M},    
\end{aligned}
\label{eq_unknowns}
\end{equation}
corresponding to $\alpha$, $\beta$, and the (independent) entries of
$\mathbf{S}_\mathcal{FF}$, $\mathbf{S}_\mathcal{FM}$, $\mathbf{S}_\mathcal{MM}$, $\mathbf{S}_\mathcal{GF}$, and $\mathbf{S}_\mathcal{GM}$ under reciprocity.
In principle, we could select a random series of $p$ known realizations of $\mathbf{b}$ and measure the corresponding $\mathbf{R}$ and $\mathbf{T}$ with a VNA and a mechanical translation stage to realize the VAA. Given the triplets $\{\mathbf{b},\mathbf{R},\mathbf{T}\}$, we could jointly estimate all $n_\mathrm{u}$ parameters via gradient descent, analogously to the procedure in~\cite{sol2023experimentally} for an RIS-parametrized end-to-end MIMO channel. However, the measurement time associated with the VAA quickly becomes prohibitively high. Indeed, for a VAA consisting of a single mechanically scanned probe, the number of required measurements would be $pN_\mathrm{G}$. Moreover, the computational burden of jointly estimating all parameters via gradient descent can be prohibitively large. The last two terms in (\ref{eq_unknowns}) scale linearly with $N_\mathrm{G}$. For a square VAA grid with $n\times n$ points, $N_\mathrm{G}=n^2$ such that both the number of unknowns associated with $\mathbf{S}_\mathcal{GF}$ and $\mathbf{S}_\mathcal{GM}$ as well as the measurement effort of a scanned-probe VAA scale quadratically with $n$. For these reasons, we aim to segment the parameter estimation problem into smaller subproblems, and to avoid gradient descent in subproblems that can be solved robustly in closed form.

Our overall strategy thus consists in first estimating all parameters involved in (\ref{eq_MNT_R}), which does not involve the VAA, and to then estimate the additional parameters involved in (\ref{eq_MNT_T}). The main challenge of estimating $\mathbf{S}_\mathcal{MM}$, i.e., the MC between the tunable lumped elements, is thereby accomplished without any VAA measurements.

Before detailing our algorithm step-by-step, we recall an important feature of (\ref{eq_MNT}). Specifically, upon changing the configuration of a single tunable lumped element, the rank of the corresponding changes of $\mathbf{R}$ and $\mathbf{T}$ is unity. This feature of the MNT model recently played an important role in achieving (i) efficient repeated MNT forward evaluations (see Sec.~\ref{sec_optimization} and~\cite{prod2023efficient}), (ii) perfect non-invasive focusing in complex media~\cite{sol2025optimal}, (iii) estimating large scattering matrices with a few-port VNA in closed form~\cite{del2025virtual,del2025virtual_3p0}, (iv) experimentally estimating MNT parameters for RIS-parametrized end-to-end MIMO channels in a segmented manner~\cite{del2025experimental,del2025ambiguity,del2025rank}, and (v) explaining the phenomenon of frozen differential scattering~\cite{delhougne2025frozen}.
Let us assume the DMA is initially in a reference configuration $\mathbf{b}_0$ and then switched to a configuration $\mathbf{b}_i$ that differs from $\mathbf{b}_0$ only regarding the $i$th meta-element's state. We define $\Delta\mathbf{T}_i\triangleq \mathbf{T}_i-\mathbf{T}_0$ and $\Delta\mathbf{R}_i\triangleq \mathbf{R}_i-\mathbf{R}_0$. Then, we perform a singular value decomposition (SVD) of $\Delta\mathbf{T}_i$ and $\Delta\mathbf{R}_i$, yielding
\begin{subequations}
\begin{equation}
    \begin{aligned}
        \Delta\mathbf{R}_i&=\mathbf{U}_{\mathrm{R},i}\,\mathbf{\Sigma}_{\mathrm{R},i}\,\mathbf{V}_{\mathrm{R},i}^\dagger         \\&=\mathbf{U}_{\mathrm{R},i}\,\mathrm{diag}([\rho_{1,i}, \dots, \rho_{N_\mathrm{F},i}])\,\mathbf{V}_{\mathrm{R},i}^\dagger,
    \end{aligned}
\end{equation}
\begin{equation}
    \begin{aligned}
        \Delta\mathbf{T}_i&=\mathbf{U}_{\mathrm{T},i}\,\mathbf{\Sigma}_{\mathrm{T},i}\,\mathbf{V}_{\mathrm{T},i}^\dagger         \\&=\mathbf{U}_{\mathrm{T},i}\,\begin{bmatrix}
\mathrm{diag}([\tau_{1,i},\dots,\tau_{\tilde N,i}]) & \mathbf{0}\\
\mathbf{0} & \mathbf{0}
\end{bmatrix}\,\mathbf{V}_{\mathrm{T},i}^\dagger,
    \end{aligned}
\end{equation}
\end{subequations}
where $\rho_{k,i}$ and $\tau_{k,i}$ denote the singular values of $\Delta\mathbf{R}_i$ and $\Delta\mathbf{T}_i$ in non-increasing order, respectively, and $\tilde{N}\triangleq\min(N_\mathrm{G},N_\mathrm{F})$. $\mathbf{U}_{\mathrm{R},i}$, $\mathbf{V}_{\mathrm{R},i}$, $\mathbf{U}_{\mathrm{T},i}$, and $\mathbf{V}_{\mathrm{T},i}$ are unitary complex-valued matrices whose columns are the left and right singular vectors of $\Delta\mathbf{R}_i$ and $\Delta\mathbf{T}_i$. We denote by $\mathbf{u}_{\mathrm{R},i}$ and $\mathbf{v}_{\mathrm{R},i}$ the first left and right singular vector of $\Delta\mathbf{R}_i$, respectively, and we denote by $\mathbf{u}_{\mathrm{T},i}$ and $\mathbf{v}_{\mathrm{T},i}$ the first left and right singular vector of $\Delta\mathbf{T}_i$. 
As detailed in Appendix~\ref{appendix_single_toggle_SVD}, the rank of $\Delta\mathbf{R}_i$ and $\Delta\mathbf{T}_i$ is unity, which implies that
\begin{enumerate}[label=(\roman*)]
    \item $\rho_{k,i} = 0 \, \forall \, k>1$,
    \item $\tau_{k,i} = 0 \, \forall \, k>1$,
    \item $\mathbf{u}_{\mathrm{R},i}$ (as well as $\mathbf{v}_{\mathrm{R},i}^*$ due to reciprocity) is collinear with $\mathbf{S}_{\mathcal{F}\mathcal{M}_i}$ if $\alpha=0$ and $\mathbf{b}_0=\mathbf{0}$,
    \item $\mathbf{u}_{\mathrm{T},i}$ is collinear with $\mathbf{S}_{\mathcal{G}\mathcal{M}_i}$ if $\alpha=0$ and $\mathbf{b}_0=\mathbf{0}$,
    \item $\mathbf{v}_{\mathrm{R},i}$ and $\mathbf{v}_{\mathrm{T},i}$ are collinear,
    \item $\mathbf{v}_{\mathrm{R},i}$ and $\mathbf{v}_{\mathrm{T},i}$ are collinear with $\mathbf{S}_{\mathcal{M}_i\mathcal{F}}^\dagger$ if $\alpha=0$ and $\mathbf{b}_0=\mathbf{0}$.
\end{enumerate}
We leverage these properties in the remainder of this paper.

\subsection{Step-by-step algorithm}
\label{subsec_stepbystep}

As mentioned, our algorithm for estimating proxy MNT parameters consists of two phases, of which only the second phase involves VAA measurements.

\textit{Phase 1}

\textit{Step 1.1}: We define $\tilde{\alpha}=0$\footnote{Our choice to define $\tilde{\alpha}=0$ should not be confused with an assumption that $\alpha=0$. The physical reflection coefficient $\alpha$ is not known to vanish, but we make the choice to assume that our proxy $\tilde{\alpha}$ vanishes. This choice reduces the number of complex-valued unknown parameters to be estimated by one. } and $\mathbf{b}_0 = \mathbf{0}$. We measure $\mathbf{R}(\mathbf{b}_0)$ and define $\tilde{\mathbf{S}}_\mathcal{FF}\triangleq \mathbf{R}(\mathbf{b}_0)$.

\textit{Step 1.2}: For each $1 \leq i \leq N_\mathrm{M}$ in turn, we measure $\mathbf{R}(\mathbf{b}_i)$, evaluate the SVD of $\Delta\mathbf{R}_i$, and identify $\mathbf{v}_{\mathrm{R},i}$ and $\mathbf{u}_{\mathrm{R},i}$. We build a reciprocal ``consensus'' vector $\mathbf{p}_i\in\mathbb{C}^{N_\mathrm{F}}$ by extracting the dominant left singular vector of their two-column stack $\left[\mathbf{u}_{\mathrm{R},i},\mathbf{v}_{\mathrm{R},i}^* \right]$.
We define $\tilde{\mathbf{S}}_{\mathcal{M}_i\mathcal{F}}\triangleq a_i \mathbf{p}_{i}^\top$, where $a_i\in\mathbb{C}$ remains to be determined in Step 1.3 below. Moreover, due to reciprocity, we define $\tilde{\mathbf{S}}_{\mathcal{F}\mathcal{M}_i} \triangleq \tilde{\mathbf{S}}_{\mathcal{M}_i\mathcal{F}}^\top = a_i \mathbf{p}_{i}$.

\textit{Step 1.3}: For $p_\mathrm{R}$ known random realizations of $\mathbf{b}$, we measure the corresponding $\mathbf{R}(\mathbf{b}^{(m)})$, where $\mathbf{b}^{(m)}$ denotes the $m$th random realization of $\mathbf{b}$. We collect all real-valued unknowns in  $\boldsymbol\theta_\mathrm{R} \in\mathbb{R}^{u_\mathrm{R}}$, where $u_\mathrm{R}=2+2N_\mathrm{M}+N_\mathrm{M}(N_\mathrm{M}+1)$. The three terms in the definition of $u_\mathrm{R}$ correspond, respectively, to $\tilde{\beta}$, the scaling factors $a_i$, and the entries of $\tilde{\mathbf{S}}_\mathcal{MM}=\tilde{\mathbf{S}}_\mathcal{MM}^\top$. 
We define the cost function to be minimized during training as
\begin{equation}
L_\mathrm{R}(\boldsymbol\theta_\mathrm{R}) = \frac{\displaystyle\sum_{m=1}^{\hat{p}_\mathrm{R}}\bigl\|\,{\mathbf R}^\mathrm{PRED}(\mathbf b^{(m)};\boldsymbol\theta_\mathrm{R})
-\mathbf R^\mathrm{MEAS}(\mathbf{b}^{(m)})\,\bigr\|_{1}}
{\displaystyle\sum_{m=1}^{\hat{p}_\mathrm{R}}\bigl\|\,\mathbf R^\mathrm{MEAS}(\mathbf{b}^{(m)})\,\bigr\|_{1}},
\label{eq9}
\end{equation}
where $\|\cdot\|_{1}$ denotes the element-wise sum of magnitudes over all complex-valued entries. The superscripts MEAS and PRED indicate, respectively, whether the channel matrix is measured or predicted based on our MNT model in (\ref{eq_MNT_R}) with parameters $\boldsymbol\theta_\mathrm{R}$.
We split our $p_\mathrm{R}$ pairs $\{\mathbf{b}^{(m)},\mathbf{R}(\mathbf{b}^{(m)})\}$ into $10$ pairs for validation and $\hat{p}_\mathrm{R} = p_\mathrm{R}-10$ pairs for training. Using the Adam optimizer with a decaying step size, we optimize $\boldsymbol\theta_\mathrm{R}$ to minimize $L_\mathrm{R}(\boldsymbol\theta_\mathrm{R})$. We monitor the validation loss, which is defined analogously to the training loss in~(\ref{eq9}) but with 10 instead of $\hat{p}_\mathrm{R}$, and stop training when it plateaus. We retain the set of parameters corresponding to the lowest validation loss.
We repeat this procedure ten times with different random initializations, and retain the set of parameters that yielded the smallest validation loss.

\textit{Phase 2}

\textit{Step 2.1}: We measure $\mathbf{T}(\mathbf{b}_0)$ and define $\tilde{\mathbf{S}}_\mathcal{GF}\triangleq \mathbf{T}(\mathbf{b}_0)$.

\textit{Step 2.2}: For each $1 \leq i \leq N_\mathrm{M}$ in turn, we measure $\mathbf{T}(\mathbf{b}_i)$, evaluate the SVD of $\Delta\mathbf{T}_i$ and identify $\mathbf{u}_{\mathrm{T},i}$. We define $\tilde{\mathbf{S}}_{\mathcal{G}\mathcal{M}_i}\triangleq h_i \mathbf{u}_{\mathrm{T},i}$, where $h_i\in\mathbb{C}$ remains to be determined in Step 2.3 below.

\textit{Step 2.3}: For $p_\mathrm{T}$ known random realizations of $\mathbf{b}$, we measure the corresponding $\mathbf{T}(\mathbf{b}^{(m)})$, where $\mathbf{b}^{(m)}$ denotes the $m$th random realization of $\mathbf{b}$. We define $\mathbf{h}=[h_1, \dots , h_{\mathrm{N}_M}]^\top\in\mathbb{C}^{N_\mathrm{M}}$ and formulate a least-squares problem to estimate $\mathbf{h}$. Specifically, we define $\tilde{\mathbf\Phi}^{(m)}=\mathrm{diag}(\tilde\beta\,\mathbf b^{(m)})$, $\tilde{\mathbf W}^{(m)}=\left(\mathbf I_{N_\mathrm M}-\tilde{\mathbf\Phi}^{(m)}\tilde{\mathbf S}_{\mathcal{MM}}\right)^{-1}\tilde{\mathbf\Phi}^{(m)}$, and $\mathbf Z^{(m)}=\tilde{\mathbf W}^{(m)}\tilde{\mathbf S}_{\mathcal{MF}} $. Then, we parametrize $\tilde{\mathbf{S}}_\mathcal{GM}$ as $\tilde{\mathbf S}_{\mathcal{GM}}=\mathbf U_\mathrm T\,\mathrm{diag}(\mathbf h)$, where $\mathbf U_\mathrm T=[\mathbf u_{\mathrm T,1},\dots,\mathbf u_{\mathrm T,N_\mathrm M}] \in\mathbb C^{N_\mathrm G\times N_\mathrm M}$. 
Substituting these newly defined variables into \eqref{eq_MNT_T} yields
\begin{equation}
\mathbf T(\mathbf b^{(m)}) = \tilde{\mathbf S}_{\mathcal{GF}} + \mathbf U_\mathrm T\,\mathrm{diag}(\mathbf h)\,\mathbf Z^{(m)} .
\end{equation}
We define $\mathbf G^{(m)} \triangleq \mathbf T(\mathbf b^{(m)})-\tilde{\mathbf S}_{\mathcal{GF}}$ and obtain
\begin{equation}
\mathbf G^{(m)} = \mathbf U_\mathrm T\,\mathrm{diag}(\mathbf h)\,\mathbf Z^{(m)} .
\label{eq_11}
\end{equation}
Next, we vectorize~(\ref{eq_11}) to obtain a linear system in $\mathbf h$:
\begin{equation}
\mathrm{vec}(\mathbf G^{(m)}) = \big(\mathbf Z^{(m)\top}\odot \mathbf U_\mathrm T\big)\,\mathbf h.
\end{equation}
With $\mathbf g^{(m)} \triangleq \mathrm{vec}\!\left(\mathbf G^{(m)}\right)\in\mathbb C^{N_\mathrm{G}N_\mathrm{F}}$ and $\mathbf A^{(m)} \triangleq \left(\mathbf Z^{(m)\top}\odot \mathbf U_\mathrm{T}\right)\in\mathbb C^{N_\mathrm{G}N_\mathrm{F}\times N_\mathrm{M}}$ we stack the $p_\mathrm{T}$ linear systems as follows:
\begin{subequations}
\begin{equation}
\mathbf g \triangleq \left[ \left(\mathbf  g^{(1)}\right)^\top, \dots, \left(\mathbf g^{(p_\mathrm{T})}\right)^\top\right]^\top \in \mathbb C^{p_\mathrm{T}N_\mathrm{G}N_\mathrm{F}},
\end{equation}
\begin{equation}
\mathbf A \triangleq \left[ \left(\mathbf A^{(1)}\right)^\top, \dots,\left(\mathbf A^{(p_\mathrm{T})}\right)^\top\right]^\top \in \mathbb C^{p_\mathrm{T}N_\mathrm{G}N_\mathrm{F}\times N_\mathrm{M}},
\end{equation}
\end{subequations}
so that $\mathbf g=\mathbf A\,\mathbf h$.
Finally, we solve for $\mathbf{h}$ as follows:
\begin{equation}
\hat{\mathbf h}=\arg\min_{\mathbf h\in\mathbb C^{N_\mathrm M}}\|\mathbf A\mathbf h-\mathbf g\|_2^2
= \mathbf A^+ \mathbf g.
\end{equation}
It follows that $\tilde{\mathbf S}_{\mathcal{GM}}=\mathbf U_\mathrm T\,\mathrm{diag}(\hat{\mathbf h})$.

\subsection{MC-unaware benchmark parameter estimation algorithm}
\label{subsec_BenchmarkParamEstimation}

To find the benchmark model parameters for $\mathbf{T}^\mathrm{noMC}$ and $\mathbf{R}^\mathrm{noMC}$, we begin by rewriting~(\ref{eq_noMC}) to group together parameters that cannot be separated unambiguously:
\begin{subequations}
\label{eq_MCunaware_regrouped}
\begin{equation}
\label{eq15}
\begin{aligned}
\mathbf{T}^\mathrm{noMC} &= {\mathbf{S}}_\mathcal{GF} +{\mathbf{S}}_\mathcal{GM} \, \mathbf{\Phi} \, {\mathbf{S}}_\mathcal{MF}
\\&={\mathbf{S}}_\mathcal{GF} +{\mathbf{S}}_\mathcal{GM} \, \mathrm{diag}(\alpha\mathbf{1}_{N_\mathrm{M}}+(\beta-\alpha)\mathbf{b}) \, {\mathbf{S}}_\mathcal{MF}
\\&= \mathbf{T}_0  + \check{\mathbf{S}}_\mathcal{GM}  \mathrm{diag}(\mathbf{b}) \, {\mathbf{S}}_\mathcal{MF}
\\&= \mathbf{T}_0  + \sum_{i=1}^{N_\mathrm{M}} b_i\check{\mathbf{S}}_{\mathcal{G}\mathcal{M}_i}   {\mathbf{S}}_{\mathcal{M}_i\mathcal{F}},
\end{aligned}
\end{equation}
\begin{equation}
\label{eq:noMC_R_grouped}
\begin{aligned}
\mathbf{R}^\mathrm{noMC}
&= \mathbf{S}_\mathcal{FF}+\mathbf{S}_\mathcal{FM}\,\mathrm{diag}(\alpha\mathbf{1}_{N_\mathrm{M}}+(\beta-\alpha)\mathbf{b})\,\mathbf{S}_\mathcal{MF}\\
&= \mathbf{R}_0 + \sum_{i=1}^{N_\mathrm{M}} b_i\,\check{\mathbf{S}}_{\mathcal{F}\mathcal{M}_i}\,\check{\mathbf{S}}_{\mathcal{F}\mathcal{M}_i}^\top.
\end{aligned}
\end{equation}
\end{subequations}
In~(\ref{eq15}), $\mathbf{T}_0 = \mathbf{S}_\mathcal{GF} + \alpha \mathbf{S}_\mathcal{GM} \, {\mathbf{S}}_\mathcal{MF}$ and $\check{\mathbf{S}}_\mathcal{GM}  = (\beta-\alpha)\,\mathbf{S}_\mathcal{GM}$.
In~(\ref{eq:noMC_R_grouped}), $\mathbf{R}_0 \triangleq \mathbf{S}_\mathcal{FF} + \alpha\,\mathbf{S}_\mathcal{FM}\mathbf{S}_\mathcal{MF}$ and
$\check{\mathbf{S}}_{\mathcal{F}\mathcal{M}_i}\triangleq \sqrt{\beta-\alpha}\,\mathbf{S}_{\mathcal{F}\mathcal{M}_i}$.

If the MC-unaware models in~(\ref{eq_MCunaware_regrouped}) accurately described the experimental reality, proxy values for all regrouped parameters could be extracted via SVDs based on one reference measurement and $N_\mathrm{M}$ measurements for single-toggle configurations. However, because of the inevitable mismatch due to MC, we extract only orientations from such SVDs while fixing scaling factors by formulating least-squares problems that we solve based on additional measurements with random DMA configurations. In addition, this choice  facilitates the comparison with our estimation of proxy MNT parameters. Technical details about the parameter estimation for our benchmark models is included in Appendix~\ref{Appendix_BenchmarkModelParameterEstimation}.

\section{Experimental Validation}
\label{sec_ExpValidation}

\subsection{DMA prototype}
\label{subsec_prototype}

\begin{figure*}
    \centering
    \includegraphics[width=2\columnwidth]{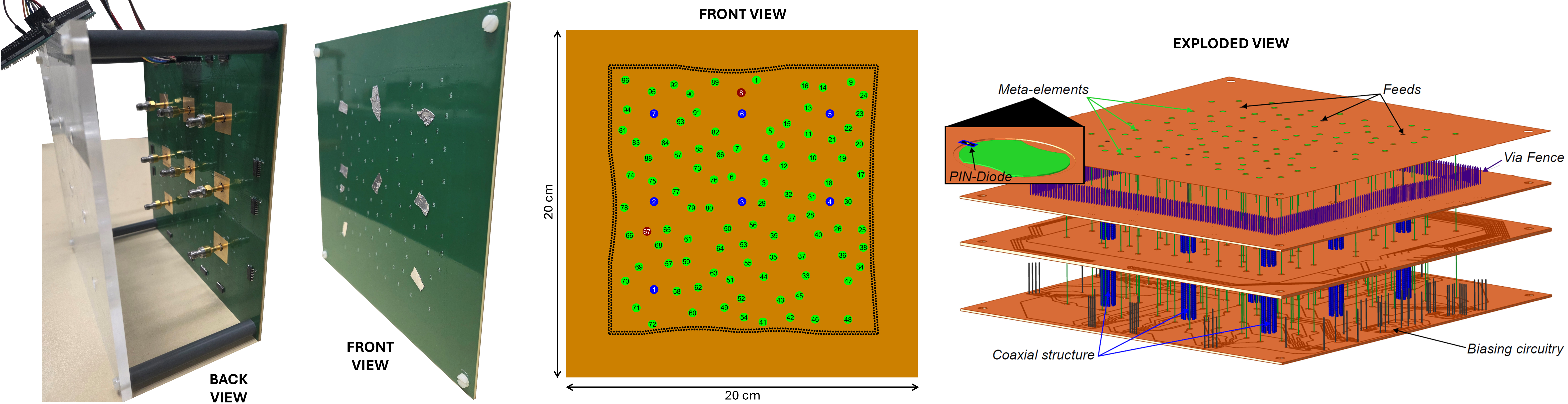}
    \caption{Photographic and schematic views of our fabricated multi-port chaotic-cavity-based DMA. The front-view layout displays the locations of the seven feeds used in our experiments (blue), the 94 functioning meta-elements (green), the two faulty meta-elements (red), and the via fence (black).}
    \label{FigDMA}
\end{figure*}

Our DMA prototype is displayed in Fig.~\ref{FigDMA} and comprises eight\footnote{As described in Sec.~\ref{subsec_setup}, we only use seven of the eight feeds in our experiment and leave the eighth feed open-circuited; it thus acts like a scatterer in the chaotic cavity.} feeds and 96 meta-elements; all feeds and meta-elements are coupled via a quasi-2D chaotic cavity. Our choice of a chaotic-cavity-based design is motivated by the recent finding that strong inter-element MC boosts the tunability of the radiated field~\cite{prod2024mutual,MCbenefitsDMA}; moreover, this design constitutes the most challenging conceivable scenario to test our algorithm for experimental proxy MNT parameter estimation.
Our DMA design closely follows the one described in~\cite{sleasman2020implementation}, except that our DMA has eight feeds while the DMA in~\cite{sleasman2020implementation} only has a single feed. Although our DMA prototype constitutes the first experimental implementation of a \textit{multi}-port chaotic-cavity-based DMA (to the best of our knowledge), the key contributions of this paper do not lie in the DMA design or fabrication but in the experimental estimation of proxy MNT parameters for the fabricated multi-port DMA prototype. In that regard, we emphasize that~\cite{sleasman2020implementation} did \textit{not} achieve model-based control of its single-port chaotic-cavity-based DMA. Instead,~\cite{sleasman2020implementation} relied on near-field scans of the radiated field corresponding to a fixed sequence of random DMA configurations in order to use the DMA for computational imaging. The absence of an accurate forward model precluded leveraging this DMA for advanced computational imaging schemes that require a model-based optimization of the utilized DMA configurations, such as task-specific end-to-end optimizations~\cite{del2020learned,qian2022noise}. Likewise, the absence of a forward model precluded efficient in-software optimizations of the DMA configuration for functionalities in wireless communications such as beamforming.\footnote{Establishing a beamforming codebook in a model-agnostic manner by sequentially optimizing the DMA configuration for different scenarios based on countless experimental measurements is prohibitively costly.}

The quasi-2D cavity of our DMA spans an area of approximately $15\times15\ \mathrm{cm}^2$ and is formed by two parallel copper layers and an irregularly shaped fence consisting of 808 vias. The irregular fence shape is chosen on purpose to break symmetries and induce wave chaos. The cavity is filled with a low-loss substrate (Rogers 4003) with a thickness of 1.52~mm.
Eight coaxial feeds (3811-40092) are connected to the cavity from the back: each feed's center pin is connected to a via which in turn is connected to an isolated circular patch on the DMA’s front surface; each feed's outer conductor is connected to the grounded cavity wall. An annular gap surrounding the circular patch on the DMA's front surface enables capacitive matching and separates it from the adjacent cavity wall. The 96 meta-elements are arranged in a pseudo-random pattern on the front side of the DMA (inside the via fence perimeter), maintaining a minimum spacing of 1.5~cm. The orientation is the same for all meta-elements. 
The meta-element design follows~\cite{yoo2016efficient}, incorporating a complementary electric-LC (cELC) resonator tuned by a PIN diode (MADP000907-14020W). Each meta-element can be switched to an ON (radiating) state or an OFF (non-radiating) state by adjusting the PIN diode's bias voltage. The state of every meta-element is controllable independently. Control over the meta-elements is achieved using 5~V~DC voltages delivered via biasing vias from twelve 8-bit shift registers (SN74AHCT595D) and four buffers (SN74ABT126D), which are positioned on the two lower copper layers (reserved for control circuitry) at the back of the DMA. Reconfigurations are executed through an Arduino microcontroller that accepts a 96-element binary vector from Python and adjusts the registers as needed. At the targeted operation frequency of 19~GHz, the size of the utilized PIN diodes (0.457~mm) is very small relative to the free-space wavelength (15.78~mm) and the wavelength in the Rogers 4003 substrate (8.38~mm). Our present paper is limited to the single-frequency regime. Insights about the instantaneous bandwidth and tunability bandwidth of our chaotic-cavity-based DMA can be found in~[Fig.~8 and Table~I,~\cite{MCbenefitsDMA}] and [Fig.~2,~\cite{yven2025end}].

As seen in the front view in Fig.~\ref{FigDMA}, we have covered the eight feed patches and two faulty meta-elements\footnote{The PIN diode is missing or not properly soldered for the two faulty meta-elements (\#8 and \#67).} with metal to ensure that leakage from the cavity to the VAA only originates from functioning meta-elements, but, to be clear, our system model does \textit{not} require this since it makes no assumption that waves only leak via functioning meta-elements from the cavity to the VAA.

\subsection{Measurement setup and procedure}
\label{subsec_setup}

\begin{figure}
    \centering
    \includegraphics[width=\columnwidth]{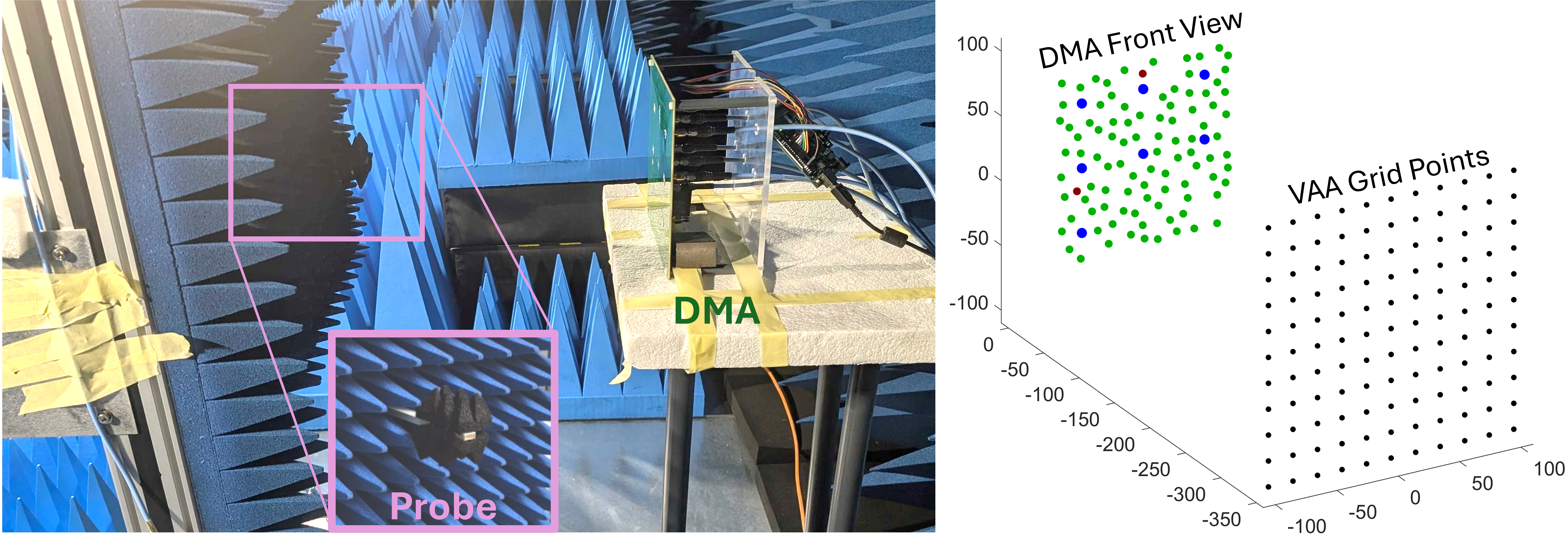}
    \caption{Photographic and schematic view of our measurement setup. Lengths are indicated in mm in the schematic.}
    \label{FigSetup}
\end{figure}

Our measurement setup is displayed in Fig.~\ref{FigSetup}. The VAA is realized based on a 2D translation stage that displaces an open-ended waveguide probe (42EWGS-A1, A-INFO INC.) in a plane parallel to the DMA at a distance of 35~cm from the DMA's front; we recall that our model makes no assumptions about the distance between DMA and VAA. An $11\times11$ regular square grid with a step size of 2~cm is scanned, implying $N_\mathrm{G}=11^2=121$. The VAA probe's orientation is aligned with the dominant polarization of the DMA's meta-element. The VAA probe is connected to one port of an eight-port VNA (two cascaded Keysight P5024B 4-port VNAs). The VNA's seven remaining ports are connected to seven DMA feeds. The remaining eighth DMA feed remains open-circuited throughout all measurements; it thus acts simply as a scatterer in the chaotic cavity. Our single-frequency measurements at 19~GHz are realized with a power of 13~dBm and an intermediate-frequency bandwidth of 500~Hz. The VNA is calibrated to include the effect of the coaxial cables connecting the VNA to the VAA probe and the feeds; thus the calibration plane is located at the connectors of the feeds and the VAA probe.

\begin{figure}
    \centering
    \includegraphics[width=\columnwidth]{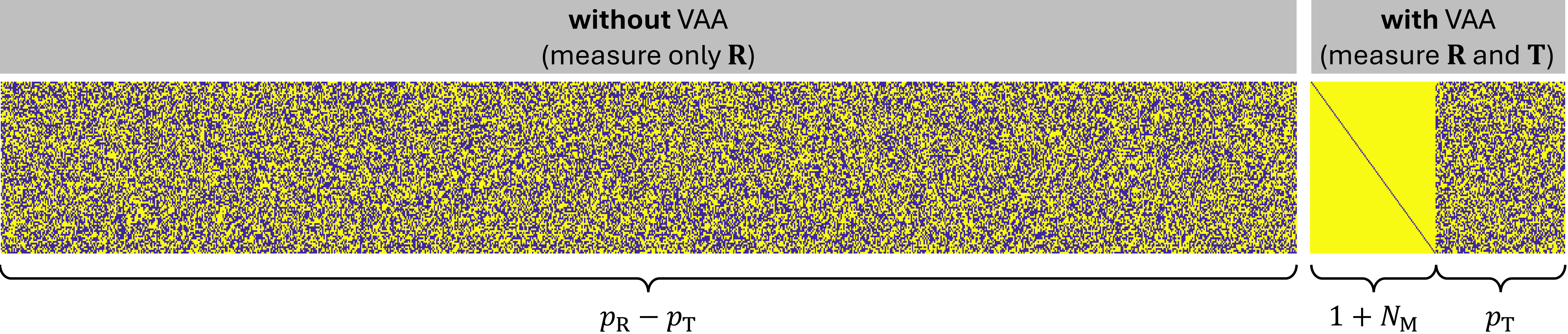}
    \caption{Sequence of measured DMA configurations. The vertical dimension corresponds to the meta-element index. Yellow and blue correspond to $b_i=0$ and $b_i=1$, respectively.}
    \label{FigProcedure}
\end{figure}

Our sequence of measurements is illustrated in Fig.~\ref{FigProcedure}, in line with the needs of the algorithm described in Sec.~\ref{sec_algorithm}. We first measure only $\mathbf{R}$ (i.e., without using the VAA) for $p_\mathrm{R}-p_\mathrm{T}$ random DMA configurations. Then, we measure both $\mathbf{R}$ and $\mathbf{T}$ (i.e., using the VAA) for a specific sequence of $1+N_\mathrm{M}$ DMA configurations followed by a sequence of $p_\mathrm{T}$ random DMA configurations. The sequence of $1+N_\mathrm{M}$ specific DMA configuration is required for Step 1.1, Step 1.2, Step 2.1, and Step 2.2. 
In our experiment, we choose $p_\mathrm{R}=1000$ and $p_\mathrm{T}=60$ by default. As explained, our algorithm is designed such that $p_\mathrm{T} \ll p_\mathrm{R}$ to minimize the number of DMA configurations measured with the VAA.
We deliberately do not treat the two faulty meta-elements differently because our algorithm can handle faulty meta-elements, as seen and discussed in Sec.~\ref{subsec_ExpParamEstim}.

Based on our measured data, we can also consider smaller values of $p_\mathrm{T}$ and $p_\mathrm{R}$ by simply ignoring a subset of our measured data. In addition, we can consider values of $N_\mathrm{F}$ that are smaller than seven by selecting a diagonal block of the measured $\mathbf{R}$ and correspondingly ignoring certain columns of the measured $\mathbf{T}$, which is equivalent to terminating the ignored feeds (among the seven measured feeds) with matched loads. In principle, it is also possible to calculate $\mathbf{R}$ and $\mathbf{T}$ for fewer used feeds for other terminations of the ignored feeds, but by choosing matched loads as terminations we ensure that the effective absorption of the cavity is independent of the number of considered feeds.

To monitor the stability of our experiment over the course of the measurements, we intermittently re-measure $\mathbf{R}$ for the same random DMA configuration. Treating the difference between the measurements of $\mathbf{R}$ under nominally identical conditions as noise, we evaluate an effective signal-to-noise ratio (SNR) to quantify the experiment's stability. Over the course of our measurements, the stability metric does not drop below 54.8~dB.

\subsection{System model sanity checks}
\label{subsec_SanityChecks}

In this subsection, we report sanity checks on the validity of our system model for our DMA prototype. Specifically, we check whether $\Delta\mathbf{R}_i$ and $\Delta\mathbf{T}_i$ are approximately of rank one (as in our preliminary work in~\cite{MultiPortDMA_EuCAP2026}), and whether $\mathbf{v}_{\mathrm{R},i}$, $\mathbf{v}_{\mathrm{T},i}$, and $\mathbf{u}_{\mathrm{R},i}^*$ are approximately collinear. Note that these sanity checks apply equally to our MNT model and our MC-unaware benchmark model.
We measure $\mathbf{R}$ and $\mathbf{T}$ for a reference configuration and all 96 single-element toggles with respect to the reference configuration. Then, we evaluate the SVDs of $\Delta\mathbf{R}_i$ and $\Delta\mathbf{T}_i$.
We define two metrics to quantify the dominance of the first singular value in $\Delta\mathbf{R}_i$ and $\Delta\mathbf{T}_i$:
\begin{subequations}
\label{eq:sanity_rankone_metrics}
\begin{align}
\gamma^{\mathrm{A}}_{\mathrm{R},i} &\triangleq \frac{\rho_{1,i}}{\rho_{2,i}},
&
\gamma^{\mathrm{A}}_{\mathrm{T},i} &\triangleq \frac{\tau_{1,i}}{\tau_{2,i}},
\label{eq:sanity_gammaA}
\\
\gamma^{\mathrm{B}}_{\mathrm{R},i} &\triangleq \frac{\rho_{1,i}^2}{\sum_{k=1}^{N_\mathrm{F}}\rho_{k,i}^2},
&
\gamma^{\mathrm{B}}_{\mathrm{T},i} &\triangleq \frac{\tau_{1,i}^2}{\sum_{k=1}^{\tilde N}\tau_{k,i}^2}.
\label{eq:sanity_gammaB}
\end{align}
\end{subequations}
Moreover, we quantify the collinearity of $\mathbf{v}_{\mathrm{R},i}$, $\mathbf{v}_{\mathrm{T},i}$, and $\mathbf{u}_{\mathrm{R},i}^*$:
\begin{subequations}
\label{eq:sanity_collinearity_metrics}
\begin{equation}
\gamma^{\mathrm{C}}_i \triangleq \left|\mathbf{v}_{\mathrm{R},i}^\dagger\mathbf{v}_{\mathrm{T},i}\right|,
\qquad
\gamma^{\mathrm{D}}_i \triangleq \left|\mathbf{v}_{\mathrm{R},i}^\dagger\mathbf{u}_{\mathrm{R},i}^*\right|.
\label{eq:sanity_gammaCD}
\end{equation}
\end{subequations}

\begin{figure}
    \centering
    \includegraphics[width=0.7\columnwidth]{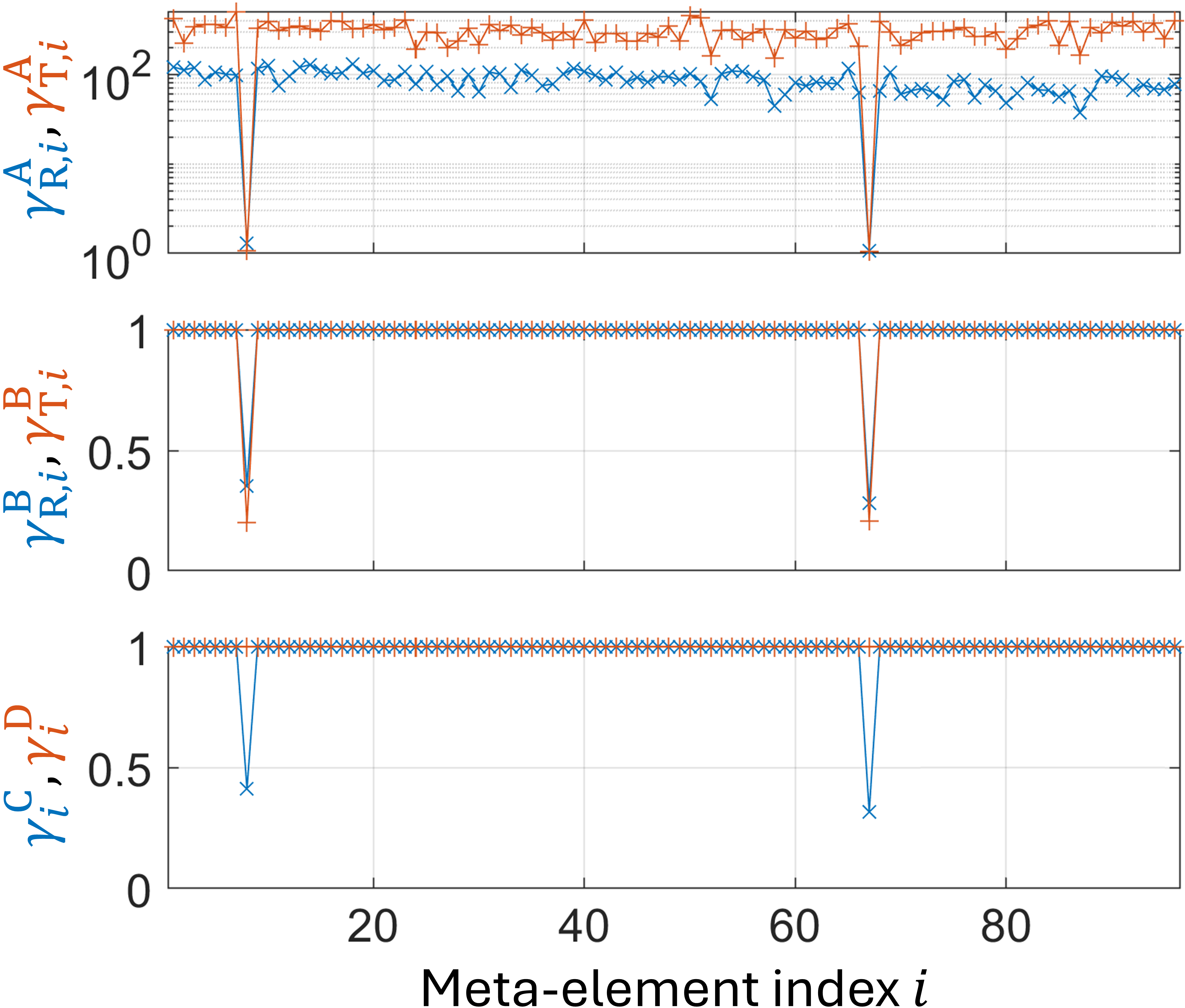}
    \caption{Sanity checks of the MNT single-element-toggle predictions.}
    \label{FigSanityChecks}
\end{figure}

The results of our sanity checks are displayed in Fig.~\ref{FigSanityChecks}. Except for $i=8$ and $i=67$ (which are the two meta-elements known to be mal-functioning), all expectations of the system model are satisfied. The first singular values dominate, as evidenced by very large values of $\gamma^\mathrm{A}_{\mathrm{R},i}$ and $\gamma^\mathrm{A}_{\mathrm{T},i}$ (which tend to infinity in the ideal model) and $\gamma^\mathrm{B}_{\mathrm{R},i}$ and $\gamma^\mathrm{B}_{\mathrm{T},i}$ being very close to unity. Moreover, the collinearity of $\mathbf{v}_{\mathrm{R},i}$, $\mathbf{v}_{\mathrm{T},i}$, and $\mathbf{u}_{\mathrm{R},i}^*$ is confirmed by $\gamma^\mathrm{C}_{i}$ and $\gamma^\mathrm{D}_{i}$ being very close to unity. For the two mal-functioning meta-elements, marked deviations from the ideal behavior are seen except for $\gamma^\mathrm{D}_i$ (because mal-functioning does not break reciprocity). Our sanity checks can thus be used as a diagnostic tool to detect faulty meta-elements.

\subsection{Model parameter estimation}
\label{subsec_ExpParamEstim}

For our prototype with $N_\mathrm{F}=7$ and $N_\mathrm{M}=96$, and our VAA with $N_\mathrm{G}=121$, the MNT system model has $n_\mathrm{u}=17{,}821$ complex-valued unknowns.

\begin{figure}
    \centering
    \includegraphics[width=0.8\columnwidth]{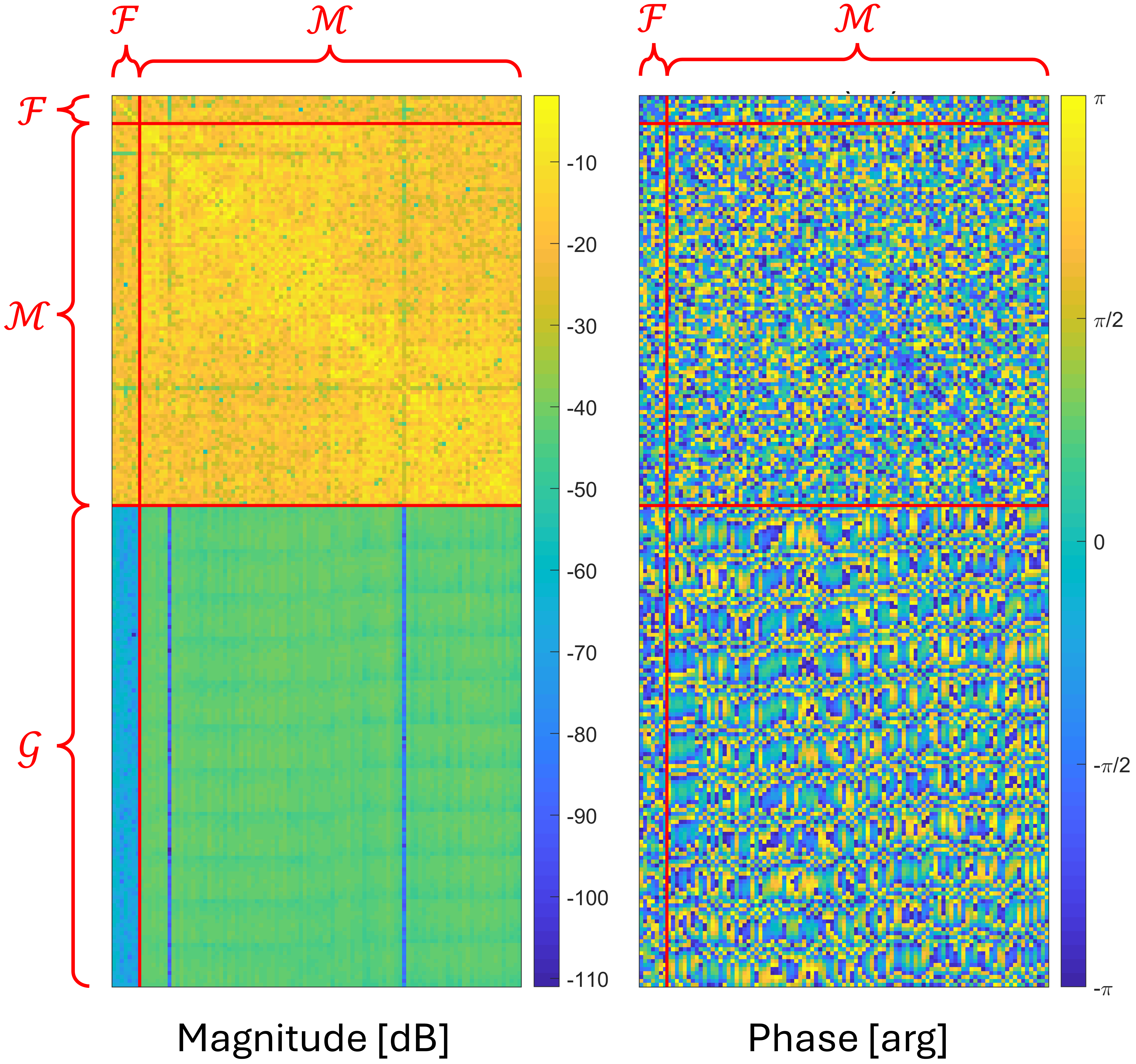}
    \caption{Example set of proxy MNT parameters estimated with $p_\mathrm{T}=60$ and $\hat{p}_\mathrm{R}=990$. The associated possible load reflection coefficients are $\tilde{\alpha}=0$ and $\tilde{\beta}=-0.3548 + 0.3001\jmath$. With reference to (\ref{eq1}), only the first two columns of the $3 \times 3$ block partition of $\tilde{\mathbf{S}}$ are shown. $\tilde{\mathbf{S}}_\mathcal{FG}$ and $\tilde{\mathbf{S}}_\mathcal{MG}$ are equal to $\tilde{\mathbf{S}}_\mathcal{GF}^\top$ and $\tilde{\mathbf{S}}_\mathcal{GM}^\top$, respectively, due to reciprocity; $\tilde{\mathbf{S}}_\mathcal{GG}$ is not estimated and plays no role in (\ref{eq_MNT}). }
    \label{FigProxyMNTparameters}
\end{figure}

We begin by visually inspecting the display in Fig.~\ref{FigProxyMNTparameters} of an example set of estimated MNT proxy parameters with $\hat{p}_\mathrm{R}=990$ and $p_\mathrm{T}=60$. We observe that our algorithm handles the two faulty meta-elements by rendering them operationally irrelevant in the sense that the predictions for $\mathbf{R}$ and $\mathbf{T}$ become effectively insensitive to the corresponding entries of $\mathbf{b}$. For instance, the columns of $\tilde{\mathbf{S}}_\mathcal{GM}$ associated with the faulty meta-elements have negligible norm, indicating negligible coupling from those elements to the VAA. Our technique thus robustly handles faulty meta-elements without requiring prior diagnostic efforts. Furthermore, we observe strong \textit{all-to-all} MC between the feeds and meta-elements, which is an expected consequence of the chaotic cavity in our DMA architecture. The coupling from the meta-elements to the VAA grid points is orders of magnitude weaker than the coupling between feeds and meta-elements within the cavity. This is consistent with free-space path loss. The direct coupling from feeds to VAA grid points is yet weaker, consistent with our mitigation of leakage paths that do not originate from functioning meta-elements.

\begin{figure}
    \centering
    \includegraphics[width=\columnwidth]{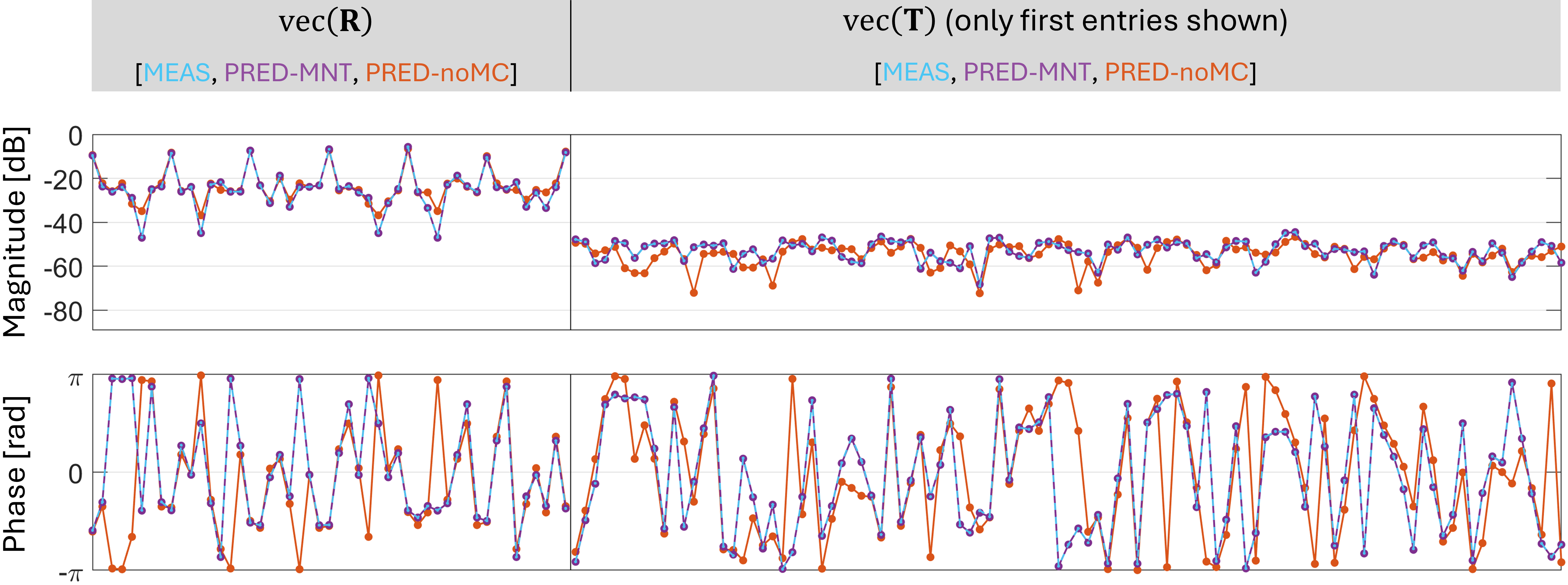}
    \caption{Visual comparison of measurement (cyan) and predictions with proxy MNT model (purple) and MC-unaware benchmark model (red) for an unseen test configuration.}
    \label{FigPredictionExample}
\end{figure}

\begin{figure*}
    \centering
    \includegraphics[width=1.5\columnwidth]{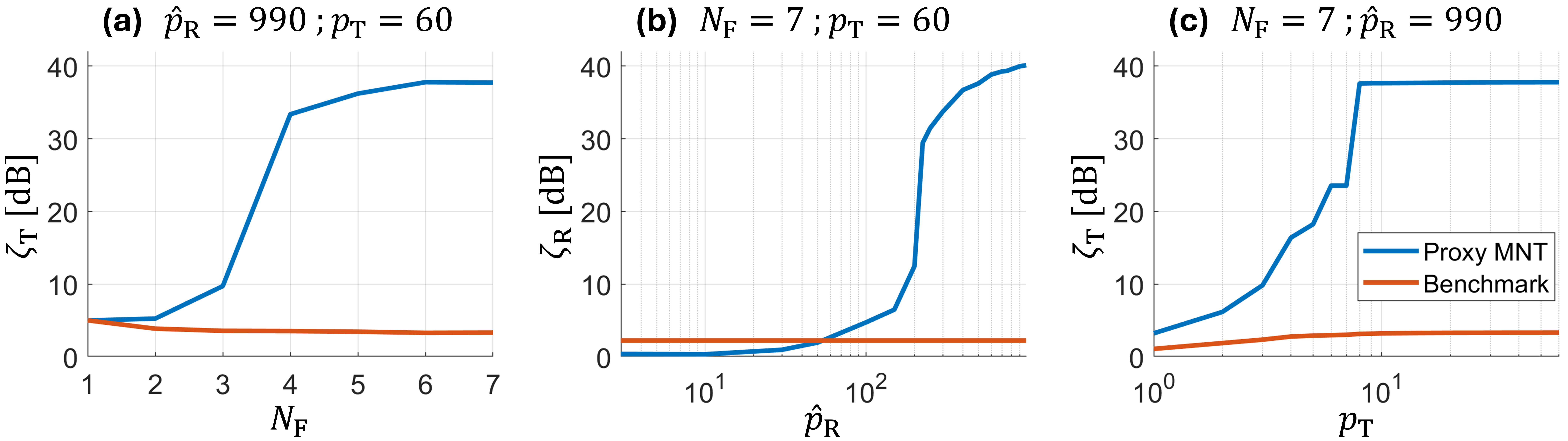}
    \caption{Dependence of model accuracies on (a) $N_\mathrm{F}$, (b) $\hat{p}_\mathrm{R}$, and (c) $p_\mathrm{T}$. }
    \label{FigAblation}
\end{figure*}

Next, we inspect the visual comparison between measurement and prediction of $\mathbf{R}$ and $\mathbf{T}$ for an unseen test configuration in Fig.~\ref{FigPredictionExample}. The proxy MNT model appears to be very accurate because no difference between the measurement and the proxy MNT model's prediction can be discerned visually. In contrast, the MC-unaware benchmark model displays clear deviations from the measurement, emphasizing the importance of MC-awareness in the case of our chaotic-cavity-based DMA that was already highlighted in~\cite{yven2025end}.

Now, we analyze the model accuracies quantitatively. Since we do not have access to the ground-truth MNT parameters, and in any case we estimate \textit{proxy} MNT parameters, we assess the accuracy of our proxy MNT model based on its ability to predict $\mathbf{R}$ and $\mathbf{T}$ for a series of $30$ random, previously unseen DMA configurations.
Specifically, we evaluate
\begin{subequations}
\label{eq:proxy_accuracy_metrics}
\begin{align}
\zeta_{\mathrm T}
&\triangleq
\left\langle\frac{\operatorname{SD}_q\!\Big(T_{\mathrm{meas},ij}^{(q)}\Big)}
     {\operatorname{SD}_q\!\Big(T_{\mathrm{meas},ij}^{(q)}-T_{\mathrm{pred},ij}^{(q)}\Big)}\right\rangle_{i,j},
\label{eq:zetaT}
\\
\zeta_{\mathrm R}
&\triangleq
\left\langle\frac{\operatorname{SD}_q\!\Big(R_{\mathrm{meas},ij}^{(q)}\Big)}
     {\operatorname{SD}_q\!\Big(R_{\mathrm{meas},ij}^{(q)}-R_{\mathrm{pred},ij}^{(q)}\Big)}\right\rangle_{i,j},
\label{eq:zetaR}
\end{align}
\end{subequations}
where $T_{\mathrm{meas},ij}^{(q)}$ denotes the $(i,j)$th entry of $\mathbf{T}$ measured for the $q$th test configuration of the DMA, and $\operatorname{SD}_q$ denotes the standard deviation across the test index $q$, and $\langle\cdot\rangle_{i,j}$ denotes the average over the two matrix indices $i$ and $j$.
The metrics $\zeta_{\mathrm T}$ and $\zeta_{\mathrm R}$ are defined analogously to an SNR, treating the mismatch between model prediction and measurement as the noise. For the proxy MNT model displayed in Fig.~\ref{FigProxyMNTparameters} and whose prediction in Fig.~\ref{FigPredictionExample} was flawless, we obtain $\zeta_\mathrm{R}^\mathrm{MNT}=40.3\ \mathrm{dB}$ and $\zeta_\mathrm{T}^\mathrm{MNT}=37.7\ \mathrm{dB}$. Meanwhile, the MC-unaware benchmark model only achieves $\zeta_\mathrm{R}^\mathrm{noMC}=2.6\ \mathrm{dB}$ and $\zeta_\mathrm{T}^\mathrm{noMC}=3.3\ \mathrm{dB}$. These values are consistent with our visual inspection of the model accuracies in Fig.~\ref{FigPredictionExample}.

So far, we used large values of $N_\mathrm{F}$, $\hat{p}_\mathrm{R}$ and $p_\mathrm{T}$. We now conduct systematic ablation studies for each of these parameters in turn to quantify how they affect the model accuracy.

We first consider $N_\mathrm{F}$, which we can reduce in integer steps down to unity ($N_\mathrm{F}=1$ corresponds to a single-feed DMA). We plot the dependence of $\zeta_\mathrm{T}$ on $N_\mathrm{F}$ in Fig.~\ref{FigAblation}(a); similar trends for $\zeta_\mathrm{R}$, as well as versions of $\zeta_\mathrm{T}$ and $\zeta_\mathrm{R}$ limited to the central feed, are not shown for the sake of conciseness. We observe in Fig.~\ref{FigAblation}(a) a clear transition between $N_\mathrm{F}=3$ and $N_\mathrm{F}=4$ where $\zeta^\mathrm{MNT}_\mathrm{T}$ jumps from 9.7~dB to 33.4~dB. For larger values of $N_\mathrm{F}$, $\zeta^\mathrm{MNT}_\mathrm{T}$ slowly converges to 37.7~dB. Meanwhile, for a single-feed DMA with $\zeta^\mathrm{MNT}_\mathrm{T}=5.0$~dB, our segmented parameter estimation is not successful. These observations suggest that even if a DMA is intended for single-feed operation, it may nonetheless be useful for the DMA to have ``auxiliary calibration feeds'' that are only used for estimating the model parameters and then open-circuited during operation. Based on MNT, a reduced version of $\mathbf{S}$ taking into account the scattering by the open-circuited feeds can be determined straightforwardly from $\mathbf{S}$. Let us imagine that the DMA's feeds are partitioned into two disjoint groups of $N_\mathrm{O}$ ``operational feeds'' (intended for use during operation) and $N_\mathrm{A}$ ``auxiliary calibration feeds'' (open-circuited during operation). We denote the corresponding sets of port indices with $\mathcal{O}$ and $\mathcal{A}$, such that $\mathcal{F}=\mathcal{O}\cup\mathcal{A}$ and $N_\mathrm{F}=N_\mathrm{O}+N_\mathrm{A}$. Then, with $\mathcal{R}=\mathcal{O}\cup\mathcal{M}\cup\mathcal{G}$ and $N_\mathrm{R}=N_\mathrm{O}+N_\mathrm{M}+N_\mathrm{G}$, the reduced scattering matrix $\mathring{\mathbf{S}}\in\mathbb{C}^{N_\mathrm{R}\times N_\mathrm{R}}$ is
\begin{equation}
    \mathring{\mathbf{S}} = \mathbf{S}_\mathcal{RR} + \mathbf{S}_\mathcal{RA} \left( \mathbf{I}_{N_\mathrm{A}} - \mathbf{S}_\mathcal{AA} \right)^{-1} \mathbf{S}_\mathcal{AR}.
\end{equation}
Then, $\mathcal{F}$ and $\mathbf{S}$ in (\ref{eq_MNT}) can be replaced by $\mathcal{O}$ and $\mathring{\mathbf{S}}$.

Next, we examine the influence of $\hat{p}_\mathrm{R}$ (with $N_\mathrm{F}=7$ and $p_\mathrm{T}=60$) in Fig.~\ref{FigAblation}(b). We observe a clear transition between $\hat{p}_\mathrm{R}=200$ and $\hat{p}_\mathrm{R}=225$ where the accuracy jumps from $\zeta^\mathrm{MNT}_\mathrm{R}=12.5\ \mathrm{dB}$ to $\zeta^\mathrm{MNT}_\mathrm{R}=29.4\ \mathrm{dB}$. For larger values of $\hat{p}_\mathrm{R}$ the accuracy still continues to rise notably, reaching $\zeta^\mathrm{MNT}_\mathrm{R}=40.3\ \mathrm{dB}$ at $\hat{p}_\mathrm{R}=990$.

Finally, we study the influence of $p_\mathrm{T}$ on $\zeta_\mathrm{T}$ in Fig.~\ref{FigAblation}(c) with $N_\mathrm{F}=7$ and $\hat{p}_\mathrm{R}=990$. We see a clear threshold of $p_\mathrm{T}=8$ beyond which further increasing $p_\mathrm{T}$ brings no appreciable improvement of $\zeta^\mathrm{MNT}_\mathrm{T}$. The $p_\mathrm{T}$ VAA measurements for random DMA configurations are used in a least-squares problem with $N_\mathrm{M}$ complex-valued unknowns in Step 2.3; each of the $p_\mathrm{T}$ measurements contributes $N_\mathrm{F}N_\mathrm{G}$ complex-valued data points. However, the condition $p_\mathrm{T}>N_\mathrm{M}/(N_\mathrm{F}N_\mathrm{G})$ is not necessarily sufficient if measurements at nearby VAA are highly correlated. This explains why the curve in Fig.~\ref{FigAblation}(c) saturates only for $p_\mathrm{T}\geq 8$ even though $N_\mathrm{M}/(N_\mathrm{F}N_\mathrm{G})<1$ in our case.

\begin{figure*}
    \centering
    \includegraphics[width=2.05\columnwidth]{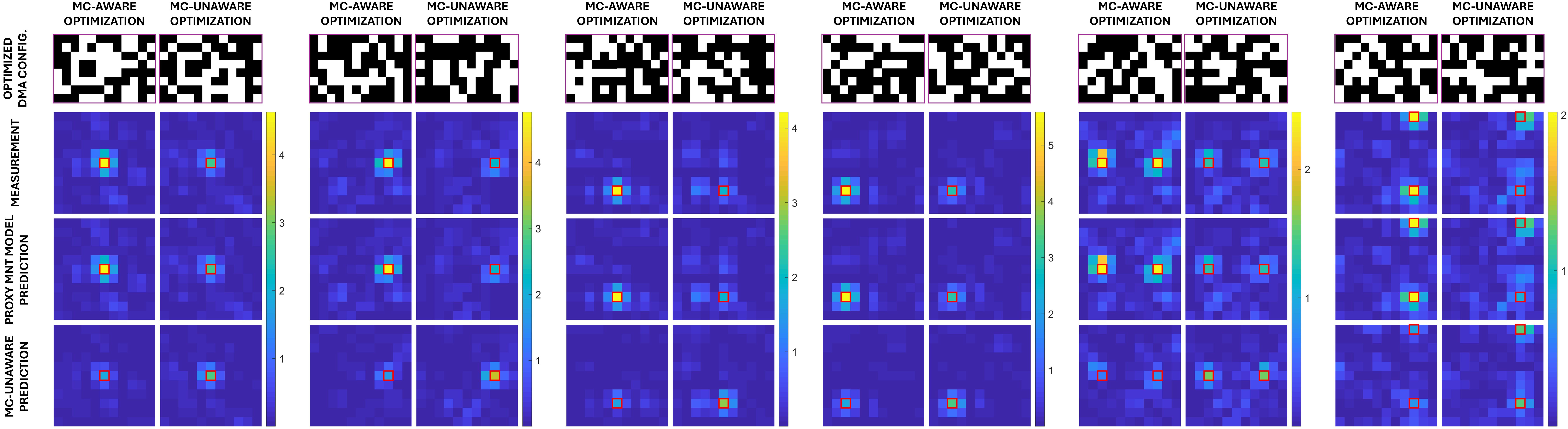}
    \caption{Examples of model-based DMA optimization for hybrid single-user and dual-user focusing. By ``hybrid'' we refer to the joint optimization of the DMA configuration $\mathbf{b}$ and the input signal $\mathbf{x}$. Red boxes indicate the targeted user location(s); four single-user and two dual-user examples are shown. In each case, the DMA configuration optimized with the MC-aware proxy MNT model is displayed in the top row on the left, and the DMA configuration optimized with the MC-unaware benchmark model is displayed in the top row on the right. (The 96-element vector is reshaped into an $8\times12$ matrix only to simplify its visualization.) The corresponding measured field intensity maps (displayed as $|\mathbf{y}_\mathrm{G}|^2 \times 10^4$) are shown in the second row. The corresponding predicted field intensity maps based on the proxy MNT model and the MC-unaware benchmark model are shown in the third and fourth rows. Each pixel in the field intensity maps corresponds to one VAA grid point (see Fig.~\ref{FigSetup}). }
    \label{FigOptim}
\end{figure*}

\section{Model-Based DMA Optimization}
\label{sec_optimization}

In this section, we demonstrate examples of model-based DMA optimization. To be clear, by ``optimization'' we refer to choosing the configuration of the DMA's individual tunable loads as opposed to optimizing the static structure of the DMA prior to fabrication. The purpose of this section is to provide further evidence for the accuracy of our proxy MNT model. Indeed, optimized DMA configurations may be outliers relative to the distribution of random DMA configurations used to evaluate the model accuracy in the previous section. Moreover, our comparison between optimizations based on the proxy MNT model and the MC-unaware benchmark model provides insight into the importance of MC awareness in DMA optimization. We limit this section to hybrid single-user and dual-user focusing, which are optimization objectives that depend only on the radiated field $\mathbf{y}_\mathrm{G}$. By ``hybrid'' we refer to the joint optimization of the DMA configuration $\mathbf{b}$ and the input signal $\mathbf{x}$. More elaborate optimization objectives, such as simultaneous focusing and jammer nulling~\cite{yven2025end} (which also depends only on $\mathbf{y}_\mathrm{G}$) or simultaneous focusing and nulling the reflected signals at the feeds~\cite{sol2023reflectionless,faul2025agile} (which depends on both $\mathbf{y}_\mathrm{G}$ and $\mathbf{y}_\mathrm{F}$), are deferred to future work. 

For hybrid single-user focusing, we maximize the focused intensity at the $i_0$th VAA grid point.
Assuming maximum-ratio beamforming across the feeds~\cite{Lo_1999_MRT}, the resulting optimization
over the DMA configuration is
\begin{equation}
\max_{\mathbf b \in \mathbb{B}^{N_\mathrm{M}}}\;
\bigl\|\mathbf{T}_{\mathcal{G}_{i_0}\mathcal{F}}(\mathbf b)\bigr\|_2^2.
\label{eq:hybrid_single_focus}
\end{equation}
For hybrid dual-user focusing, we maximize the minimum focused intensity at the $i_1$th and $i_2$th VAA grid points:
\begin{equation}
\max_{\mathbf b \in \mathbb{B}^{N_\mathrm{M}}}\;
\max_{\|\mathbf x\|_2=1}\;
\min\!\Big(
\bigl|\mathbf{T}_{\mathcal{G}_{i_1}\mathcal{F}}(\mathbf b)\,\mathbf x\bigr|^2,\;
\bigl|\mathbf{T}_{\mathcal{G}_{i_2}\mathcal{F}}(\mathbf b)\,\mathbf x\bigr|^2
\Big).
\end{equation}
To identify an optimized DMA configuration for a given optimization objective, we use a coordinate descent algorithm similar to [Algorithm~1,~\cite{MCbenefitsDMA}]; we leverage the Woodbury identity to efficiently compute the updated inverse after each single-element toggle~\cite{prod2023efficient}, avoiding a full matrix inversion at every coordinate step.
In the case of dual-user focusing, the optimization of $\mathbf{x}$ is treated as an inner problem. For each candidate $\mathbf{b}$, our code computes $\mathbf T(\mathbf b)$ and then optimizes $\mathbf x$ by restricting it to $\mathrm{span}\!\left\{\mathbf T_{\mathcal G_{i_1}\mathcal F}(\mathbf b)^{\ast},\,\mathbf T_{\mathcal G_{i_2}\mathcal F}(\mathbf b)^{\ast}\right\}$ and grid-searching a two-parameter unit-norm parametrization to maximize
$\min\!\left(\left|\mathbf T_{\mathcal G_{i_1}\mathcal F}(\mathbf b)\mathbf x\right|^2,\left|\mathbf T_{\mathcal G_{i_2}\mathcal F}(\mathbf b)\mathbf x\right|^2\right)$.

Representative results are displayed in Fig.~\ref{FigOptim}. For each optimization objective, we identify one optimized configuration based on our proxy MNT model and one based on the MC-unaware benchmark model. Then, we measure $\mathbf{T}$ for the two optimized configurations and obtain the corresponding predictions of $\mathbf{T}$ based on the two models. The displayed intensity maps in Fig.~\ref{FigOptim} demonstrate flawless agreement between the measured radiated fields and their predictions based on the proxy MNT model in all cases; meanwhile, the MC-unaware benchmark's predictions display notable inaccuracies. Moreover, the focusing performance achieved with the DMA configurations optimized based on the MC-unaware benchmark model is notably inferior in all cases. These results demonstrate the importance of MC-awareness in optimizing our DMA prototype for single-user and dual-user focusing.

\section{Conclusion}
\label{sec_Conclusion}

To summarize, we have experimentally estimated an accurate proxy MNT model for a multi-port DMA with strong MC between all feeds and meta-elements. Our proxy MNT model accurately predicts both the DMA's radiated field and the field reflected at the DMA's feeds. An MC-unaware benchmark model is orders of magnitude less accurate. Our proxy MNT parameter estimation is segmented into multiple steps to minimize the number of DMA configurations measured with the VAA. Initially, we estimate the coupling between the feeds, the coupling between the feeds and the meta-elements, as well as the coupling between the meta-elements; we combine closed-form and gradient-descent steps to ensure efficiency and robustness of our approach. Then, we estimate the coupling between the meta-elements and VAA probe points in closed form. We observe that using at least four feeds notably improves the accuracy of the proxy MNT model parameter estimation, motivating the inclusion of ``auxiliary calibration feeds'' in DMAs that are ultimately intended for single-feed operation. We further observe rather sharp transitions from low to high model accuracy as a function of the number of measured DMA configurations in various algorithmic segments, indicating the minimum number of measurements to ensure the well-posedness of the problem.

Our MNT model formulation applies to any DMA architecture whose reconfigurability relies on tunable lumped elements. In particular, our presented technique can thus  be applied directly to microstrip-based DMA architectures~\cite{sleasman2015dynamic,shlezinger2021dynamic}, 3D-chaotic-cavity-based DMA architectures~\cite{sleasman2016microwave}, as well as BD-DMA architectures~\cite{prod2025beyond}. Our MNT model is also directly compatible with an end-to-end system model for a wireless DMA-based network, where the VAA ports are simply replaced by user ports. In such wireless systems, the estimated parameters can be refreshed at different time scales: Phase 1 can be conducted once before operation, while Phase 2 is repeated once per coherence time. 

Looking forward, we anticipate extensions of our presented technique to include both polarizations of the radiated field and to express the radiated field in other bases such as spherical harmonics. Moreover, we foresee generalizations of our technique to non-reciprocal DMAs (see related work in~\cite{del2025virtual_3p0,del2025virtual_3p1}) and to operating with purely non-coherent measurements (see related work in~\cite{sol2023experimentally,del2025virtual,del2025virtual_2p0,del2025virtual_3p1}). Finally, the availability of an accurate forward model enables the integration of the DMA into end-to-end optimized system pipelines, e.g., for computational imaging~\cite{del2020learned,qian2022noise}.

\appendices

\section{Single-Toggle SVD Properties}
\label{appendix_single_toggle_SVD}

With $\mathbf{W}\triangleq (\mathbf{I}_{N_\mathrm{M}}-\mathbf{\Phi}\mathbf{S}_{\mathcal{MM}})^{-1}\mathbf{\Phi}$, we can rewrite \eqref{eq_MNT} as
\begin{subequations}
\label{eq:RT_W}
\begin{align}
\mathbf{R} &= \mathbf{S}_{\mathcal{FF}} + \mathbf{S}_{\mathcal{FM}}\,\mathbf{W}\,\mathbf{S}_{\mathcal{MF}}, \\
\mathbf{T} &= \mathbf{S}_{\mathcal{GF}} + \mathbf{S}_{\mathcal{GM}}\,\mathbf{W}\,\mathbf{S}_{\mathcal{MF}} .
\end{align}
\end{subequations}
Consider a baseline DMA configuration (characterized by $\mathbf{\Phi}$) as well as a perturbed DMA configuration
(characterized by $\mathbf{\Phi}'$) differing only in the state of the $i$th meta-element. Since $\mathbf{\Phi}$ is diagonal,
we can write $\mathbf{\Phi}'=\mathbf{\Phi}+\delta\,\mathbf{e}_i\mathbf{e}_i^\top$, where $\delta \triangleq r_i^\prime-r_i \in\mathbb{C}$ and $\mathbf{e}_i$ denotes the $i$th canonical basis vector. Let us now define $\mathbf{B}\triangleq \mathbf{I}_{N_\mathrm{M}}-\mathbf{\Phi}\,\mathbf{S}_{\mathcal{MM}}$ and $\mathbf{B}^\prime\triangleq \mathbf{I}_{N_\mathrm{M}}-\mathbf{\Phi}^\prime\,\mathbf{S}_{\mathcal{MM}}
= \mathbf{B}-\delta\,\mathbf{e}_i(\mathbf{e}_i^\top\,\mathbf{S}_{\mathcal{MM}})$.
By the Sherman-Morrison identity~\cite{sherman1950adjustment}, we obtain
\begin{equation}
\begin{aligned}
\Delta\mathbf{W} \triangleq \mathbf{W}^\prime-\mathbf{W} = \nu\,\mathbf{D}\,\mathbf{e}_i\, \mathbf{e}_i^\top\,\mathbf{F},
\end{aligned}
\label{eq:DeltaW}
\end{equation}
where $\nu \triangleq \delta \left[1-\delta\,\mathbf{e}_i^\top\, \mathbf{S}_{\mathcal{MM}}\,\mathbf{D}\,\mathbf{e}_i\right]^{-1}\in\mathbb{C}$, $\mathbf{D} = (\mathbf{I}_{N_\mathrm{M}}-\mathbf{\Phi}\mathbf{S}_{\mathcal{MM}})^{-1}\in\mathbb{C}^{N_\mathrm{M}\times N_\mathrm{M}}$ and
$\mathbf{F} = (\mathbf{I}_{N_\mathrm{M}}-\mathbf{S}_{\mathcal{MM}}\mathbf{\Phi})^{-1}\in\mathbb{C}^{N_\mathrm{M}\times N_\mathrm{M}}$.

Inserting \eqref{eq:DeltaW} into \eqref{eq:RT_W} yields
\begin{subequations}
\label{eq:DeltaRDeltaT_outer}
\begin{align}
\Delta\mathbf{R}_i
&= \mathbf{S}_{\mathcal{FM}}\,\Delta\mathbf{W}\,\mathbf{S}_{\mathcal{MF}}
= \big(\mathbf{S}_{\mathcal{FM}}\,\mathbf{D}\,\mathbf{e}_i\big)\,\nu\,
\big(\mathbf{e}_i^\top\,\mathbf{F}\,\mathbf{S}_{\mathcal{MF}}\big), \\
\Delta\mathbf{T}_i
&= \mathbf{S}_{\mathcal{GM}}\,\Delta\mathbf{W}\,\mathbf{S}_{\mathcal{MF}}
= \big(\mathbf{S}_{\mathcal{GM}}\,\mathbf{D}\,\mathbf{e}_i\big)\,\nu\,
\big(\mathbf{e}_i^\top\,\mathbf{F}\,\mathbf{S}_{\mathcal{MF}}\big).
\end{align}
\end{subequations}
Both $\Delta\mathbf{R}_i$ and $\Delta\mathbf{T}_i$ are seen to be outer products of a column vector and a row vector in (\ref{eq:DeltaRDeltaT_outer}); thus, their rank is unity.
Moreover, \eqref{eq:DeltaRDeltaT_outer} shows that $\Delta\mathbf{R}_i$ and $\Delta\mathbf{T}_i$ share the same right factor
$\mathbf{e}_i^\top\,\mathbf{F}\,\mathbf{S}_{\mathcal{MF}}$,
implying that the dominant right singular vector of $\Delta\mathbf{R}_i$  and  the dominant right singular vector of $\Delta\mathbf{T}_i$ are collinear.

Assuming that the base configuration is $\mathbf{b}=\mathbf{0}$ and $\alpha=0$, it follows that the baseline DMA configuration is characterized by $\mathbf{\Phi}=\mathbf{0}$ and the perturbed DMA configuration is characterized by $\mathbf{\Phi}' = \beta\,\mathbf{e}_i\,\mathbf{e}_i^\top$ (with $\beta\neq 0$).
It further follows that $\mathbf{W}=\mathbf{0}$ and
\begin{equation}
\mathbf{W}'=(\mathbf{I}_{N_\mathrm{M}}-\mathbf{\Phi}^\prime\,\mathbf{S}_{\mathcal{MM}})^{-1}\,\mathbf{\Phi}^\prime
= \frac{\beta}{1-\beta\,\mathbf{S}_{{\mathcal{M}_i}{\mathcal{M}_i}}}\,\mathbf{e}_i\,\mathbf{e}_i^\top .
\end{equation}
Inserting this expression into \eqref{eq:RT_W} yields
\begin{subequations}
\begin{align}
\Delta\mathbf{R}_i &= \kappa_i\,\mathbf{S}_{\mathcal{F}\mathcal{M}_i}\,\mathbf{S}_{\mathcal{M}_i\mathcal{F}},\\
\Delta\mathbf{T}_i &= \kappa_i\,\mathbf{S}_{\mathcal{G}\mathcal{M}_i}\,\mathbf{S}_{\mathcal{M}_i\mathcal{F}},
\end{align}
\end{subequations}
with $\kappa_i=\beta\,\left[1-\beta\,\mathbf{S}_{{\mathcal{M}_i}{\mathcal{M}_i}}\right]^{-1}$. Hence, $\Delta\mathbf{R}_i$ and $\Delta\mathbf{T}_i$ are rank-one outer products whose left factors are $\mathbf{S}_{\mathcal{F}\mathcal{M}_i}$ and $\mathbf{S}_{\mathcal{G}\mathcal{M}_i}$, respectively, and whose common right factor is $\mathbf{S}_{\mathcal{M}_i\mathcal{F}}$. Therefore, the dominant left singular vectors of $\Delta\mathbf{R}_i$ and $\Delta\mathbf{T}_i$ are collinear with $\mathbf{S}_{\mathcal{F}\mathcal{M}_i}$ and $\mathbf{S}_{\mathcal{G}\mathcal{M}_i}$, respectively, and their dominant right singular vectors are collinear with each other and with $\mathbf{S}_{\mathcal{M}_i\mathcal{F}}^\dagger$.

\section{MC-Unaware Benchmark Model \\Parameter Estimation}
\label{Appendix_BenchmarkModelParameterEstimation}

To estimate a set of proxy parameters (denoted by an overhead bar) for the MC-unaware benchmark model in~(\ref{eq15}), we proceed as follows:

\textit{Step B.T.1}: We measure $\mathbf{T}(\mathbf{b}_0)$, where $\mathbf{b}_0=\mathbf{0}$ and define $\bar{\mathbf{T}}_0\triangleq \mathbf{T}(\mathbf{b}_0)$.

\textit{Step B.T.2}: For each $1 \leq i \leq N_\mathrm{M}$ in turn, we measure $\mathbf{T}(\mathbf{b}_i)$, evaluate the SVD of $\Delta\mathbf{T}_i$ and identify $\mathbf{u}_{\mathrm{T},i}$ and $\mathbf{v}_{\mathrm{T},i}$. We define $\bar{\check{\mathbf{S}}}_{\mathcal{G}\mathcal{M}_i}\triangleq \phi_i \mathbf{u}_{\mathrm{T},i}$, where $\phi_i\in\mathbb{C}$ remains to be determined in Step B.T.3 below, and $\bar{\mathbf{S}}_{\mathcal{M}_i\mathcal{F}}\triangleq  \mathbf{v}_{\mathrm{T},i}^\dagger$.

\textit{Step B.T.3}: For $p_\mathrm{T}$ known random realizations of $\mathbf{b}$, we measure the corresponding $\mathbf{T}(\mathbf{b}^{(m)})$, where $\mathbf{b}^{(m)}$ denotes the $m$th realization. We define $\Delta\mathbf{T}^{(m)} \triangleq \mathbf{T}(\mathbf{b}^{(m)})-\bar{\mathbf{T}}_{0}\approx\sum_{i=1}^{N_\mathrm{M}} b^{(m)}_i\,\phi_i\,\mathbf{u}_{\mathrm{T},i}\mathbf{v}_{\mathrm{T},i}^\dagger$. Vectorizing yields a linear system of equations for the unknown parameters $\boldsymbol\phi\triangleq[\phi_1,\dots,\phi_{N_\mathrm{M}}]^\top\in\mathbb{C}^{N_\mathrm{M}}$:
\begin{equation}
\mathbf{t}^{(m)} \triangleq \mathrm{vec}\!\left(\Delta\mathbf{T}^{(m)}\right)
\approx \mathbf{Q}^{(m)}\,\boldsymbol\phi,
\end{equation}
where $\mathbf{Q}^{(m)} \triangleq \left[ \mathbf{q}^{(m)}_1, \dots, \mathbf{q}^{(m)}_{N_\mathrm{M}}\right]\in\mathbb{C}^{N_\mathrm{G}N_\mathrm{F}\times N_\mathrm{M}}$ and $ \mathbf{q}^{(m)}_i \triangleq b^{(m)}_i\,\mathrm{vec}\!\left(\mathbf{u}_{\mathrm{T},i}\mathbf{v}_{\mathrm{T},i}^\dagger\right) $.
We stack the $p_\mathrm{T}$ equations as follows:
\begin{subequations}
\label{eq_17}
\begin{align}
\mathbf{t} &\triangleq \left[ \left(\mathbf{t}^{(1)}\right)^\top , \dots , \left( \mathbf{t}^{(p_\mathrm{T})} \right)^\top \right]^\top 
\in\mathbb{C}^{p_\mathrm{T}N_\mathrm{G}N_\mathrm{F}},\\
\mathbf{Q} &\triangleq \left[\left(\mathbf{Q}^{(1)}\right)^\top, \dots, \left(\mathbf{Q}^{(p_\mathrm{T})}\right)^\top\right]^\top \in\mathbb{C}^{p_\mathrm{T}N_\mathrm{G}N_\mathrm{F}\times N_\mathrm{M}},
\end{align}
\end{subequations}
so that $\mathbf{t}\approx \mathbf{Q}\boldsymbol\phi$. Finally, we solve
\begin{equation}
\hat{\boldsymbol\phi}
=\arg\min_{\boldsymbol\phi\in\mathbb{C}^{N_\mathrm{M}}}\|\mathbf{Q}\boldsymbol\phi-\mathbf{t}\|_2^2
=\mathbf{Q}^+\,\mathbf{t}.
\end{equation}

To estimate the proxy parameters (denoted by an overhead bar) for the MC-unaware benchmark model in~(\ref{eq:noMC_R_grouped}), we proceed analogously, as detailed in the following:

\textit{Step B.R.1}: We measure $\mathbf{R}(\mathbf{b}_0)$ with $\mathbf{b}_0=\mathbf{0}$ and define $\bar{\mathbf{R}}_0\triangleq \mathbf{R}(\mathbf{b}_0)$.

\textit{Step B.R.2}: For each $1\leq i\leq N_\mathrm{M}$ in turn, we measure $\mathbf{R}(\mathbf{b}_i)$, form $\Delta\mathbf{R}_i\triangleq \mathbf{R}(\mathbf{b}_i)-\bar{\mathbf{R}}_0$, and compute the dominant singular vectors of $\Delta\mathbf{R}_i$. We build a reciprocal ``consensus'' vector $\mathbf{p}_i\in\mathbb{C}^{N_\mathrm{F}}$ by extracting the dominant left singular vector of their two-column stack $\left[\mathbf{u}_{\mathrm{R},i},\mathbf{v}_{\mathrm{R},i}^* \right]$. We define $\bar{\check{\mathbf{S}}}_{\mathcal{F}\mathcal{M}_i} \triangleq \sqrt{\omega_i} \mathbf{p}_i$, where $\omega_i\in\mathbb{C}$ remains to be determined in Step B.R.3; the square-root ambiguity is irrelevant because only $\omega_i$ but not $\sqrt{\omega_i}$ is used in (\ref{eq:noMC_R_grouped}).

\textit{Step B.R.3}: For $\hat{p}_\mathrm{R}$ known random DMA configurations $\mathbf{b}^{(m)}$, we measure $\mathbf{R}(\mathbf{b}^{(m)})$ and evaluate $\Delta\mathbf{R}^{(m)}\triangleq \mathbf{R}(\mathbf{b}^{(m)})-\bar{\mathbf{R}}_0$. With $\mathbf{P}_i \triangleq \mathbf{p}_i\,\mathbf{p}_i^\top$, we have $\Delta\mathbf{R}^{(m)} \approx \sum_{i=1}^{N_\mathrm{M}} b^{(m)}_i\,\omega_i\,\mathbf{P}_i$. 
We define $\mathbf{d}^{(m)}\triangleq \mathrm{vec}\big((\Delta\mathbf{R}^{(m)})_{\mathcal{U}}\big)\in\mathbb{C}^{N_\mathcal{U}}$ and
$\mathbf{c}_i\triangleq \mathrm{vec}\!\big((\mathbf{P}_i)_{\mathcal{U}}\big)\in\mathbb{C}^{N_\mathcal{U}}$, where $\mathcal{U}$ denotes the index set of the upper-triangular entries and $N_\mathcal{U}=N_\mathrm{F}(N_\mathrm{F}+1)/2$. Then,
\begin{equation}
\mathbf{d}^{(m)} \approx \mathbf{C}^{(m)}\,\boldsymbol\omega,
\end{equation}
where $\mathbf{C}^{(m)} \triangleq \big[\,b_1^{(m)}\mathbf{c}_1,\dots,b_{N_\mathrm{M}}^{(m)}\mathbf{c}_{N_\mathrm{M}}\,\big]$.
Stacking the $\hat{p}_\mathrm{R}$ systems analogously to (\ref{eq_17}) yields $\mathbf{d}\approx \mathbf{C}\,\boldsymbol\omega$, which we solve in least squares as
\begin{equation}
\hat{\boldsymbol\omega}
=\arg\min_{\boldsymbol\omega\in\mathbb{C}^{N_\mathrm{M}}}\|\mathbf{C}\boldsymbol\omega-\mathbf{d}\|_2^2
=\mathbf{C}^+\,\mathbf{d}.
\end{equation}

\bibliographystyle{IEEEtran}


\end{document}